\documentclass[12pt,cite,epsfig]{article}
\usepackage{times,epsfig}
\newcommand{\nc}{\newcommand}
\nc{\bb}{\bibitem}
\nc{\be}{\begin{equation}}
\nc{\ee}{\end{equation}}
\nc{\pa}{\partial}
\nc{\parsym} {\stackrel{\leftrightarrow}{\pa}}
\nc{\ra}{\rightarrow}
\nc{\la}{\leftarrow}
\nc{\etp}{{\eta^\prime}}
\nc{\omg}{\omega}
\nc{\ggam}{\gamma \gamma}
\nc{\gam}{\gamma }
\nc{\bea}{\begin{eqnarray}}
\nc{\eea}{\end{eqnarray}}
\nc{\beas}{\begin{eqnarray*}}
\nc{\eeas}{\end{eqnarray*}}
\nc{\non}{\nonumber}

\def\hhhu{\rule[-3.mm]{0.mm}{11.mm}}
\def\hhhv{\rule[-3.mm]{0.mm}{9.mm}}

\begin{document}
\begin{titlepage}
\vbox{~~~ \\
                                   \null \hfill LPNHE 2009--06\\

\title{A Global Treatment Of VMD Physics Up To The $\phi$~:\\
II.~~ $\tau$ Decay  and  Hadronic Contributions To $g-2$
   }
\author{
M.~Benayoun, P.~David, L.~ DelBuono, O. Leitner \\
\small{ LPNHE Paris VI/VII, IN2P3/CNRS, F-75252 Paris, France }\\
}
\date{\today}
\maketitle
\begin{abstract}
Relying on the Hidden Local Symmetry (HLS) model equipped with a mechanism breaking
the U(3)/SU(3)/SU(2) symmetries and generating 
a dynamical vector meson mixing, it has been shown
that a global
fit successfully describes the cross sections for the $e^+ e^- \ra \pi^+ \pi^-$,
 $e^+ e^- \ra (\pi^0/\eta) \gamma$ and $e^+ e^- \ra \pi^0 \pi^+ \pi^-$ annihilation
 channels. One extends this global fit in order to include also the dipion spectra
 from the $\tau$ decay, taking into account all reported information on their statistical and
 systematic errors. A model accounting for lineshape distortions of
 the $\rho^\pm$ spectrum relative to $\rho^0$ is also examined when analyzing the $\tau$ 
 data behavior within the global fit framework. One shows that a successful account for
 $e^+ e^-$ annihilation data and $\tau$ spectra can be simultaneously reached.
  Then, issues related with non--perturbative hadronic contributions to the muon $g-2$
 are examined in details. It is shown that all $e^+ e^-$ data considered together allow for
 improved and motivated estimates for the $a_\mu(\pi^+ \pi^-)$, the $\pi^+ \pi^-$  loop 
 contribution to the muon $g-2$~; for instance,  integrated between 0.630 and 0.958 GeV,
 we find $a_\mu(\pi^+ \pi^-)= 359.62 \pm 1.62$ (in units of $10^{-10}$), a 40\% improvement of
 the current uncertainty.
 The effects of the various $\tau$ samples in the context of a global fit procedure 
 leads to  conclude that different
 lineshape distortions are revealed by the ALEPH, BELLE and CLEO data samples.  
 Relying on global 
 fits to the data quoted above, one also provides  motivated estimates of the $ \pi^+ \pi^-$, 
  $\pi^0\gamma$, $\eta \gamma$  and $\pi^0 \pi^+ \pi^-$ contributions
 to $a_\mu$ up to 1 GeV with the smallest possible uncertainties. 
 These estimates are based on various global fit configurations, each
 yielding a good probability.

\end{abstract}
}
\end{titlepage}

\section{Introduction}
\indent \indent 
It has been proved in \cite{ExtMod1} that the scope of 
the HLS model \cite{HLSRef,FKTUY}, suitably broken \cite{taupaper}, can be extended in order
to include annihilation processes like $e^+e^- \ra (\pi^0/\eta) \gamma$, 
$e^+e^- \ra \pi^+ \pi^-\pi^0$ and decay spectra like $\eta/\eta^\prime \ra \pi^+ \pi^- \gamma$,
beside the $e^+e^- \ra \pi^+ \pi^-$ cross section.
Actually, it also includes  $e^+e^- \ra K^+ K^-$ and $e^+e^- \ra K^0 \overline{K}^0$
annihilations which have to be examined separately as they raise known specific
problems \cite{BGPter}. Actually, most VMD physics up to the $\phi$ meson mass is covered 
by our extended model \cite{ExtMod1}, except for channels involving scalar mesons
or channels where higher mass vector mesons could have a significant influence \cite{omgpi}
as, seemingly, $e^+e^- \ra \omg \pi$.  However, all the $e^+e^-$ annihilations channels examined 
in \cite{ExtMod1} are reasonably well described   up to the $\phi$ mass region
by the model presented in \cite{ExtMod1}. 
 
 The issue is now to examine how the Extended (HLS) Model
performs while including other processes  like the $\tau^\pm \ra \pi^\pm \pi^0 \nu$ 
decay which is in the scope of the HLS model. This leads us to report on the results  
of global fits using the existing $\tau$ dipion spectra  beside the $e^+e^- $ data 
extensively discussed in \cite{ExtMod1}.  The same issue was partly addressed\footnote{We considered together
with $e^+e^- \ra \pi^+ \pi^-$ cross sections, the pion form factor from ALEPH
and the dipion spectrum {\it lineshape} from CLEO. We were, thus, less sensitive
to  issues related with the absolute scale of the $\tau$ spectra.}  in our
former \cite{taupaper}. 
The energy range of our model is limited approximately by the $\phi$ meson mass~; meanwhile,
as far as issues like the muon $g-2$ value are concerned, this is an energy region  
where a reliable model can address some questions in a novel way.

Indeed, an interesting outcome of such a global fit is the estimate it provides
for various hadronic contributions to the muon $g-2$. This covers the $\pi^+ \pi^-$
loop contribution, but also those from the  $(\pi^0/\eta) \gamma$ and 
$\pi^+ \pi^-\pi^0$ final states. The improvements following from using simultaneously
$\tau$ data and $e^+ e^-$ data as well as the consequences of having a reliable global 
model describing all VMD physics up to the $\phi$ meson mass are interesting issues.
 Indeed, the underlying unified (HLS) framework of our model correlates several 
 decay channels  because of their common underlying physics and phenomenological studies
 indicate that these physics correlations are well accepted by the data \cite{ExtMod1}. 
 Stated otherwise,
 one can examine in great details several consequences of accounting for 
$\tau$ decays and $e^+ e^-$ annihilations within a consistent framework.

This turns also out to readdress the long--standing problem
of the discrepancy between the BNL measurement \cite{BNL} of $g-2$ and the predictions
based on $e^+ e^-$ annihilations and $\tau$ spectra as reported in the literature
\cite{DavierPrevious1,DavierPrevious2,DavierPrevious3,Davier2003,Eidelman,Fred09}.
A quite recent study \cite{DavierHoecker} tends to lessen the disagreement
between these two kinds of predictions, but not to resorb it.

Our present study is based on all the $e^+e^- $ data  sets used in \cite{ExtMod1}
and on the published $\tau$ spectra. These are 
 the dipion mass spectra collected  some time ago by ALEPH \cite{Aleph} and 
CLEO \cite{Cleo}~; a valuable data set collected by the BELLE Collaboration,
with a statistics of several million  events,
has been made recently available \cite{Belle}.

The paper is organized as follows. In Section \ref{FitTau}, we describe the model
for the dipion spectrum in the $\tau^\pm \ra \pi^\pm \pi^0 \nu$ decay. The
vector meson mixing produced by Isospin Breaking (IB) \cite{ExtMod1,taupaper}
is complemented in a simple manner with another IB mechanism allowing 
lineshape distortions of the $\rho^\pm$ meson mass spectrum compared with $\rho^0$.
In this Section, we also list the $\tau$ data sets and outline
how they intervene in the global fit procedure. 

In Section \ref{FitProcess}, we first emphasize the correlation between
pure IB shape distortion parameters and the absolute scale of $\tau$ spectra.
Then, the consistency of the dipion spectra from $\tau$  decay 
-- not affected by vector meson mixing -- with all $e^+ e^-$ annihilation channels 
is investigated. The behavior of each $\tau$ data set  -- the ALEPH \cite{Aleph}, CLEO 
\cite{Cleo} samples and the recently issued BELLE\cite{Belle} data sample -- are examined 
under various kinds of fit conditions. 
It is shown that the ALEPH spectrum can be well described with simple and intuitive
IB lineshape distortions compared to $e^+ e^-$, whereas this does not work well
with BELLE and CLEO spectra. The best way to account for these is rather a rescaling
of the absolute scale of their spectrum. We argue that this could point towards
a more complicated  IB lineshape distortion model than ours.
One shows, nevertheless, that a satisfactory simultaneous account of all $e^+ e^-$ 
annihilation data and the available dipion spectra in the  $\tau$  decay can be reached, 
under quite reasonable conditions.

In Section \ref{gMoins2}, we focus on
the non--perturbative hadronic contributions to the muon $g-2$, especially
the $\pi^+ \pi^-$ one.  The results provided by the various $e^+ e^- \ra \pi^+ \pi^-$
data sets are examined and the effects of global fits involving  the  $e^+e^- \ra (\pi^0/\eta) \gamma$  
and $e^+e^- \ra \pi^+ \pi^-\pi^0$ cross sections is shown. The effects of
including the $\tau$ dipion spectra within the fitted data sets is examined in
full details. It is also shown that the KLOE  data set  \cite{KLOE_ISR1} for
$e^+ e^- \ra \pi^+ \pi^-$ does not lead to some inconsistency but, rather,
allows for improved uncertainties at the expense of a worse fit probability. 

We also conclude on the most likely value of several contributions to the muon
$g-2$ following from a global fit to a large sample of $e^+ e^-$  and $\tau$ data.
The uncertainties yielded look much improved with respect to usual.

Finally, Section \ref{FinalConclusion} is devoted to a summary of our conclusions.
In particular, one emphasizes using the various $\tau$ spectra in order to
provide -- or improve -- theoretical predictions for the muon $g-2$, taking
into account the difficulty to model lineshape distortions in a way accepted
simultaneously by the ALEPH, BELLE and CLEO data sets.

In the present paper, which is the second part of a study started in \cite{ExtMod1},
one does not discuss the properties or the results of the fits to
the $e^+ e^-$ data in isolation. These have been discussed at length in  \cite{ExtMod1}~;
we also refer the reader to that paper for the details of our model. 
All notations have been carefully chosen in order to ensure full consistency with \cite{ExtMod1}.
Finally, the present work supersedes and improves large parts of our former study \cite{taupaper}.
We also correct here for a (minor) computer code error which affected the treatment of
the sample--to--sample correlated uncertainties  in the data sets from
\cite{CMD2-1995corr,CMD2-1998-1,SND-1998}~; this is, indeed, important in order to provide
reliable uncertainties to our $g-2$ estimates.

\section{Including $\tau^\pm \ra \pi^\pm \pi^0 \nu$ Data}
\label{FitTau}
\indent \indent 
The difference between $e^+ e^-$ and $\tau$ based estimates of the hadronic contribution
to the  muon $g-2$ is an important issue. Indeed, accounting for isospin symmetry breaking
effects, the  $\tau$ dipion spectra{\footnote{Each normalized to the world average
branching ratio  Br$(\tau \ra \pi \pi \nu)$, highly influenced by the ALEPH measurement \cite{Aleph}.}} 
provide  predictions for the hadronic contribution
which makes the expected value of $g-2$ close to its experimental measurement \cite{BNL}.
 Instead, all theoretical estimates based on $e^+ e^-$ data deviate by
 more than 3 $\sigma$. Comprehensive discussions of this issue can be found in  
\cite{DavierPrevious1,DavierPrevious2,DavierPrevious3} and more 
recently in \cite{Fred09}. Summaries can also be found in
\cite{Davier2003,Eidelman}, for instance. A quite recent reanalysis of this discrepancy
\cite{DavierHoecker} concludes to a smaller disagreement between $\tau$ and $e^+e^-$ 
based approaches (about $2 ~\sigma$)~; consequently, the newly proposed $\tau$ based estimate 
moves farther from  the BNL measurement. However, even if reduced, the mismatch between
$e^+ e^-$ and $\tau$ based estimates of the hadronic contribution to $g-2$ survives.
 
It was shown in  \cite{taupaper} that an appropriate account of 
isospin symmetry breaking (IB), including its effects on the ($\rho,~\omg,~\phi$) mixing,
certainly  solves a part of the reported discrepancy between $e^+ e^-$ and $\tau$ spectra.
However, the IB vector mixing defined there and recalled in \cite{ExtMod1} does not
exhaust all effects of IB. In this paper, we examine more deeply
than in \cite{taupaper} the effects of IB shape distortions and their connection
with absolute scale issues. In order to examine this kind of IB,  one needs 
a data sample where the $\rho^0$ ($e^+e^-$ annihilation) and the 
$\rho^\pm$ ( $\tau$ decay) spectra are simultaneously present. 
 
 The problem of the hadronic contributions to the muon $g-2$ was not addressed 
 in \cite{taupaper}. This issue is examined here in a wider context 
by revisiting the consistency pattern of the $\tau^\pm \ra \pi^\pm \pi^0 \nu$ data 
on one hand, and the much larger data set on the $e^+e^- \ra \pi^+ \pi^-$,
$e^+e^- \ra (\pi^0/\eta) \gamma$ and $e^+e^- \ra \pi^+ \pi^-\pi^0$ annihilation 
channels on the other hand. This is allowed by having extended the model presented 
in \cite{taupaper} in such a way that anomalous and non--anomalous channels 
are implemented within the unified framework presented in \cite{ExtMod1}.

Most part of the $e^+e^-\ra \pi^+ \pi^-$ data sets  has been commented on  in
\cite{taupaper}~; here, we only remind how the sample--to--sample correlated 
part of the systematic uncertainties should be treated, as this plays an important role
in estimating the uncertainty on the muon $g-2$. All other $e^+e^-$ annihilation
channels have been considered in details in our recent \cite{ExtMod1}. Because of the poor
probability of the best fit to the KLOE data \cite{KLOE_ISR1} already 
commented on in \cite{ExtMod1},
the corresponding  data sample is not included systematically in the 
set of  $e^+e^- \ra \pi^+ \pi^-$ data samples considered~; however, its  effects will be commented 
upon at the appropriate places. Finally,
in order to fit the parameters of our $\rho,~\omg,~\phi$ mixing scheme \cite{taupaper}, 
one still uses a subset
of  9  radiative decay width data which have been taken  from the latest issue of the
Review of Particle Properties \cite{RPP2008} and are given explicitly in \cite{ExtMod1}.

\subsection{The Model For $\tau^\pm \ra \pi^\pm \pi^0 \nu$ Decay}
\indent \indent Our model for the pion form factor in $\tau$ decay 
coincides exactly with the formulae given in \cite{taupaper}~: 
\be
\displaystyle
F_\pi^\tau(s) = \left [
(1-\frac{a}{2}) - \frac{ag}{2} F_\rho^\tau(s) \frac{1}{D_\rho(s)}
\right]
\label{Nobrk0}
\ee
with~:
\be
\left \{
\begin{array}{lll}
\displaystyle F_\rho^\tau(s) =f_\rho^\tau - \Pi_{W}(s)~~~,~~~f_\rho^\tau=ag f_\pi^2\\[0.5cm]
\displaystyle D_\rho(s)=s-m^2-\Pi_{\rho \rho}(s)
\end{array}
\right .
\label{Nobrk1}
\ee
where $\Pi_{W}(s)$ accounts for the loop corrections to the $\rho^\pm - W^\mp$
transition amplitude $f_\rho^\tau$  and $\Pi_{\rho \rho}(s)$ is the $\rho^\pm$ self--mass.
Both loops are such that they vanish at $s=0$.
$g$  denotes, as usual \cite{HLSRef,ExtMod1},
the universal vector coupling and $m^2=a g^2 f_\pi^2$ is the
$\rho$ meson mass squared as it occurs in the HLS Lagrangian. $a$ is the 
standard HLS parameter
expected close to 2  \cite{HLSRef,taupaper,ExtMod1}.

Beside the  mixing of vector mesons produced by breaking Isospin Symmetry,  
Reference \cite{taupaper} 
examined the possibility of having a mass difference beween the neutral and charged 
$\rho$ mesons. Here, we also allow for a mass squared difference  between neutral 
and charged $\rho$ mesons -- denoted resp. $m^2$ and $m^2 + \delta m^2$. Additionally,
we also allow for a coupling difference of these mesons, resp. $g$
and $g^\prime = g + \delta g$. The $\rho^\pm - W^\mp$ transition
amplitude should be modified correspondingly \cite{taupaper}, as will be reminded
shortly. These two parameters correspond within our model to allowing mass and width
differences between the charged and neutral $\rho$ mesons, as commonly done
in other studies \cite{DavierHoecker,Ghozzi}.

\subsubsection{The Pion Form Factor In the $\tau$ Decay}
\label{ffth}
\indent \indent  With the IB modifications just defined,
the pion form factor has to be slightly modified compared with Eq. (\ref{Nobrk0}). 
It can be written~:
\be
\displaystyle
F_\pi^\tau(s) = \left [
(1-\frac{a}{2}) - F_\rho^\tau(s) g_{\rho \pi \pi}^\prime \frac{1}{D_\rho(s)}
\right]
\label{Model1}
\ee
 where $g_{\rho \pi \pi}^\prime=ag^\prime/2=a[g + \delta g]/2$.
 The other ingredients are modified, compared with 
 Eqs. (\ref{Nobrk0}) and (\ref{Nobrk1}), and become~:
\be
\left \{
\begin{array}{lll}
F_\rho^\tau(s) =f_\rho^{ \prime \tau} - \Pi_{W}(s)\\[0.5cm]
D_\rho(s)=s-m^2 -\delta m^2 -\Pi_{\rho \rho}^\prime(s)\\[0.5cm]
\displaystyle f_\rho^{\prime \tau} =  f_\rho^\tau + \delta f_\rho^\tau~~~,  
~~~\delta f_\rho^\tau = \frac{\delta m^2}{g^\prime} -\frac{f_\rho^\tau \delta g}{g^\prime}
\end{array}
\right .
\label{Model2}
\ee
where $f_\rho^\tau =a g  f_\pi^2$ is the $\rho- W$ transition
amplitude,  $D_\rho(s)$ is the inverse $\rho^\pm$ propagator and 
$\Pi_{\rho \rho}^\prime(s)$ is the charged $\rho$ self--mass. With the
$\delta f_\rho^\tau$ term in the last Eq. (\ref{Model2}), $F_\pi^\tau(0)=1$ 
is identically fulfilled. In \cite{taupaper}, we assumed $\delta g =0$.

The (modified)  $F_\rho^\tau(s)$ is the $W-\rho$ transition
amplitude with its loop corrections. In terms of the
pion $\ell_\pi(s)$ and kaon  $\ell_{K}(s)$ amputated loops, one has the following 
expressions~:
\be
\left \{
\begin{array}{lll}
\displaystyle \Pi_{W}(s)=g_{\rho \pi \pi}^\prime
\left [ (1-\frac{a}{2}) \ell_\pi(s) + \frac{1}{2 z_A^2}(z_A-\frac{a}{2}) \ell_{K}(s) 
\right ]+ P_W(s) \\[0.5cm]
\displaystyle \Pi_{\rho \rho}^\prime(s) = [g_{\rho \pi \pi}^\prime]^2  \left [ 
\ell_\pi(s) + \frac{1}{2 z_A^2} \ell_{K}(s)~
\right ] + P_\rho(s)
\end{array}
\right .
\label{Model3}
\ee
where $z_A=[f_K/f_\pi]^2$ is the standard SU(3) breaking parameter in the BKY breaking scheme
\cite{BKY,Heath1}, while $P_W(s)$ and $P_\rho(s)$ are subtraction polynomials with real coefficients
to be fixed by external conditions. 

One could look for a motivated way, like the BKY mechanism \cite{BKY},
 able to generate this kind of IB distortion effects. 
The proposed modifications look, however, reasonable and correspond to the usual way
of introducing mass and width differences in other studies.
This mechanism will be referred to as IB shape distortion and, if numerically relevant, 
may complement the IB vector mixing \cite{taupaper,ExtMod1}.

We have checked that one can safely identify
$\ell_\pi(s)$ and $\ell_{K}(s)$ -- both being charged--neutral meson loops -- occuring in these 
expressions with the amputated $\pi^+ \pi^-$ and $K^+ K^-$ loops 
appearing in $e^+ e^-$ annihilations \cite{taupaper}.

In order to reduce the number of free parameters in the global fit procedure,
we still identify (as in \cite{taupaper}) the subtraction polynomial for
 $\Pi_W(s)$ with those for its partner in $e^+ e^-$ annihilation
(see Section 5 in \cite{ExtMod1}). On the other hand, as one can neglect pseudoscalar meson mass
differences in loop calculations, one also identifies the charged $\rho$ self--mass
$\Pi_{\rho \rho}^\prime(s)$ with its neutral $\rho$ partner -- up to the $\delta g$ effect --
as reminded in Section 5 of \cite{ExtMod1}.

Finally, in order to fit $\tau$ data, one has to correct for specific isospin symmetry 
breaking effects.
For this purpose, short range \cite{Marciano} ($S_{EW}=1.0235$) and long range 
\cite{Cirigliano1,Cirigliano2,Cirigliano3} ($G_{EM}(s)$) radiative corrections 
have to be considered. While comparing with experimental data, the quantity
in Eq. (\ref{Model1}) has to be modified  to~:
\be
F_\pi^\tau(s) \Longrightarrow \left [ S_{EW} G_{EM}(s) \right ]^{1/2} F_\pi^\tau(s)
\label{Model5}
\ee

As the standard HLS model does not go beyond the lowest lying vector meson nonet, we cannot
fit the whole  dipion $\tau$ decay spectrum. We chose to stop around the $\phi$ mass
\cite{ExtMod1} as, up to this energy, higher mass vector mesons seem  to have a very
limited influence within the channels examined in \cite{ExtMod1} and in the present
work.

\subsubsection{Useful Expressions For The Dipion Partial Width in the $\tau$ Decay}
\indent \indent 
The dipion partial width in the decay of the $\tau$ lepton
can be written \cite{taupaper}~:
  \be
\displaystyle \frac{d\Gamma_{\pi \pi}(s)}{ds} = \displaystyle
\frac{|V_{ud}|^2 G_F^2}{64 \pi^3 m_\tau^3} |F_\pi(s)|^2 
G_0(s) + {\cal O}(\epsilon^2) 
\label{FF1}
\ee
\noindent with~:
 \be
 \left \{
\begin{array}{lll}
G_0(s) &= \displaystyle \frac{4}{3} \frac{(m_\tau^2-s)^2(m_\tau^2+2 s)}{s^{3/2}} Q_\pi^3\\[0.5cm]
 Q_\pi &= \displaystyle \frac{\sqrt{[s-(m_{\pi^0}+m_{\pi^+})^2][s-(m_{\pi^0}-m_{\pi^+})^2]}}{2\sqrt{s}}
\end{array}
\right .
\label{FF2}
\ee
\noindent where $\epsilon =(m_{\pi^0}^2 -m_{\pi^+}^2)/m_{\pi^+}^2\simeq -0.06455$. The terms of
order $\epsilon^2$ --  which manifestly break Isospin Symmetry -- are negligible.
On the other hand, one obviously has~:
 \be
\displaystyle \frac{1}{\Gamma_{\pi \pi}} \frac{d\Gamma_{\pi \pi}(s)}{ds} = 
\frac{1}{N}\frac{dN(s)}{ds}
\label{FF2b}
\ee
where $\Gamma_{\pi \pi}$ is the (integrated) $\pi \pi$ partial width in the $\tau$ decay~;
$1/N dN(s)/ds$ is the normalized spectrum of yields over the accessible dipion invariant 
mass range{\footnote{Of course, the total number of pion pairs is defined by 
$N =\int [dN(s)/ds] ~ds$.}}.  While referring to $\tau$ normalized spectra in the
following, we always understand this quantity.

Using Eqs. (\ref{FF1}) and (\ref{FF2b}) together with the customary
expression \cite{RPP2008} for the the $\tau \ra e \nu_\tau \nu_e $ partial width,
one can derive~: 
\be
\displaystyle
|F_\pi(s)|^2= \frac{2 m_\tau^8}{|V_{ud}|^2 (m_\tau^2-s)^2 (m_\tau^2+2 s) }
\frac{1}{\beta_-^3} \frac{{\cal B}_{\pi \pi}}{{\cal B}_e} \frac{1}{N}\frac{dN(s)}{ds}
\label{FF3}
\ee
\noindent which is the standard expression used by experimentalists to reconstruct the
pion form factor from experimental data  \cite{Cleo,Belle}. In this expression 
$\beta_-$ is the pion velocity in the dipion rest frame,
${\cal B}_{\pi \pi}$ and ${\cal B}_e$ are  the branching ratios of the
$\tau$ decays to resp. $\pi \pi \nu_\tau$ and to $e \nu_\tau \nu_e$. 

Eq. (\ref{FF2b}) can also be written~:
 \be
\displaystyle \frac{1}{\Gamma_{\tau}} \frac{d\Gamma_{\pi \pi}(s)}{ds} = 
{\cal B}_{\pi \pi} \frac{1}{N}\frac{dN(s)}{ds}
\label{FF4}
\ee
where $\Gamma_{\tau}$ denotes the full $\tau$ width. The theoretical
expression for $d\Gamma_{\pi \pi}/ds$ on the left--hand side is given
by Eq. (\ref{FF1}) and by
 $|F_\pi(s)|^2$ as following from Subsection \ref{ffth}  above~;
the additional factors shown by Eq. (\ref{Model5}) are be understood. Finally,
the numerical value for $\Gamma_{\tau}$ -- not accessible to our model -- is
derived  from the measured lifetime \cite{RPP2008} and  ${\cal B}_{\pi \pi}$
is numerically provided by each experiment with various uncertainties.  
Eq. (\ref{FF4}) is the main tool in the present analysis.

\subsubsection{Absolute Normalization of the Dipion $\tau$ Spectrum}
\label{tauscale1}
\indent \indent 
As clear from Eqs. (\ref{Model5}) and (\ref{FF1}), the absolute normalization
of the theoretical dipion partial width spectrum is determined by the product 
$G_F^2 |V_{ud}|^2 S_{EW} G_{EM}(s)$. Correspondingly, the absolute normalization
of the experimental spectrum on the right-hand side of Eq.(\ref{FF4}) is
determined by the branching ratio ${\cal B}_{\pi \pi}$. 

Less obvious analytically, but numerically important, is the role played
by the universal vector coupling $g$ and the transition amplitude $f_\rho^{\prime \tau}$ 
in providing the theoretical normalization of the
dipion spectrum. Indeed, as $a$ is found numerically close to 2,
Eqs. (\ref{Model1}) and (\ref{Model2}) show that the absolute magnitude
of the dipion spectrum is proportional to the product squared of the
$\rho  - W$ and $\rho  \pi \pi$ amplitudes. Therefore, actually, non--vanishing 
$\delta g$ and $\delta m^2$ influence both the lineshape and the absolute
normalization of the $\tau$ spectrum.

Moreover, one cannot exclude some other 
mechanism breaking CVC by modifying essentially the absolute
normalization of the $\tau$ spectra~; therefore, a correction factor $(1+\eta_{CVC})$
may enter  Eq. (\ref{FF4}) and can be fitted.  
Related with this, one should note a recent BaBar 
measurement  about the $\tau - \mu -e$ universality. 
BaBar reports\footnote{We thanks W. M. Morse
to have drawn our attention on this paper.} \cite{BaBar09}
$g_\mu/g_e= 1.0036 \pm 0.0020$ as expected, while  $g_\tau/g_\mu= 0.985 \pm 0.005$
exhibits a $3~\sigma$ departure from 1. If confirmed, this may indicate
a possible CVC violation in the $\tau$ sector\footnote{This BaBar result
contradicts a former measurement from ALEPH \cite{Aleph} which was
consistent with lepton universality.} affecting only the absolute
scale of the $\tau$ spectrum. On the other hand, an experimental bias
on the ${\cal B}_{\pi \pi}$ branching ratio cannot be excluded and
could play in the same direction. 

Even if very close to standard approaches, the IB lineshape distortions
have been introduced here in a very simplified manner.  One can easily think
of a more complicated structure of these than  the one we inferred.

\subsection{Dealing With The Fitted Data Sets}
\indent \indent Besides the  data sets provided\footnote{As in our 
\cite{taupaper}, we discard  the OPAL data 
set \cite{Opal}.} by the ALEPH \cite{Aleph} and CLEO \cite{Cleo} 
Collaborations, a new sample has been recently made available by the  
BELLE Collaboration\cite{Belle}.  These are the $\tau$ data sets
which will be examined in conjunction with the whole set of $e^+ e^-$ 
data samples already considered in \cite{ExtMod1}. We remind that a subset of 9 vector meson
decay partial widths is also used, corresponding to decay modes not related with the
annihilation data considered in our fit procedure~; these are numerically extracted
from \cite{RPP2008}.

\subsubsection{The $\tau$ Input To The Fit Procedure}
\label{taudata}
\indent \indent 
In the present study, we  submit to fit the experimental spectra as shown in the right--hand side
of  Eq. (\ref{FF4}). In order to remain consistent,
we use for each experiment its own published branching ratio measurement and
not the world average branching ratio.

As for the CLEO data, our input is their published spectrum \cite{Cleo} for $1/N dN(s)/ds$
normalized to their latest updated branching ratio measurement\cite{CleoBr,RPP2008}, 
${\cal B}_{\pi \pi}= (25.36 \pm 0.44) \%$. This Collaboration also claims an
uncertainty on the absolute energy scale \cite{Cleo,CleoPriv} of about 0.9 MeV.
However, in our former analysis  \cite{taupaper}, no such uncertainty showed up
significatively. Anticipating somewhat on our present 
analysis, we confirm its effective consistency with zero and, therefore,  
discard this freedom from now on.

Concerning the ALEPH data, we use directly the last update of the 
${\cal B}_{\pi \pi}/N dN(s)/ds$ spectrum  \cite{Aleph}. The corresponding
branching fraction,  ${\cal B}_{\pi \pi}= (25.471\pm 0.097 \pm 0.085)\%$,
is the most precise among the published measurements. The 
uncertainties will be added in quadrature (0.127\%).

For the BELLE data \cite{Belle}, we  have been provided \cite{HisakiPriv} with 
all information concerning the pion form factor spectrum\footnote{ Normalized 
to the world average 
branching ratio ${\rm Br}(\tau \ra \pi^+ \pi^- \nu)$, four times more precise
than the BELLE own measurement, and slightly shifted.}, its covariance matrix for statistical 
errors and its systematics. The systematics have been added in quadrature to the statistical 
error covariance matrix. The BELLE $1/N dN(s)/ds$ spectrum data are published as such \cite{Belle}~;
its error covariance matrix  can be derived from the corresponding information provided
for  the pion form factor, using simple algebra. 
As stated above, we have  submitted to fit the BELLE
${\cal B}_{\pi \pi}/N dN(s)/ds$ spectrum normalized to the BELLE branching ratio  \cite{Belle} 
${\cal B}_{\pi \pi}= (25.34 \pm 0.39) \%$.

The uncertainty provided by the branching ratio error is clearly a scale uncertainty and
a bin--to--bin correlated error~; this should be treated as reminded in Section 
6 of \cite{ExtMod1}. This turns out to define the (partial) $\chi^2$ for each of the ALEPH, BELLE and CLEO 
data sets by \cite{ExtMod1}~:
\be
\chi^2_{Exp} = [(1+\lambda_{Exp} ) m_i - f(s_i)] [(1+\lambda_{Exp} ) m_j - f(s_j)] V^{-1}_{ij}
+\left[ \frac{\lambda_{Exp}}{\eta_{Exp}} \right ]^2
\label{FF6}
\ee
having defined, for each experiment, the measurements $m_i$ as the central value 
for the branching ratio times $1/N dN(s_i)/ds$ and $V$ being the full error
covariance matrix. $f(s_i)$ should be understood as the left--hand side of Eq.
(\ref{FF4}) computed at the appropriate energy point. 

For each experiment, $\lambda$ is a scale parameter to
be fitted and $\eta$ is the ratio of the branching ratio uncertainty to its central 
value. The second term in this $\chi^2$  is the standard way to account for a scale uncertainty.
We have $\eta_{CLEO}=1.74\%$, $\eta_{ALEPH}=0.51\%$ and $\eta_{BELLE}=1.53\%$. 

With this input to the fit procedure, the  ALEPH, BELLE and CLEO data sets are
clearly treated on the same footing. 
As emphasized in \cite{ExtMod1}, if for some experiment the ratio $\lambda_{fit}/\eta_{Exp}$
is small enough (typically not greater than $\simeq 1 \div 2$), one can neglect this scale correction
and use the standard $\chi^2$ expression in the minimization procedure, with the replacement 
$V_{ij} \Rightarrow V_{ij} +\eta_{Exp}^2 m_i m_j $. Otherwise, one may consider we are
faced with some missing variance and keep Eq. (\ref{FF6}) as it stands.

 In a previous study \cite{taupaper}, we limited ourselves to considering only the $\tau$ data
 points up to 0.9 GeV in order to avoid at most effects of higher mass vector mesons.
 In the present work, however, preliminary studies using the BELLE and CLEO 
 data\footnote{For instance, fitting the  BELLE and CLEO spectrum lineshapes up to 1. GeV
 does not reveal worse fit quality than when stopping the fit at 0.9 GeV. Higher mass vector
 meson influence in this region, if any, is thus found small enough to be absorbed
 by the other fit parameters.} 
 samples lead us to push this upper energy limit  up to 1.0 GeV. Indeed, as in
 our $e^+ e^-$ fit studies \cite{ExtMod1}, the influence of higher mass vector mesons
 seems negligible all along  the energy region from threshold to 
  1.0 GeV -- and  even slightly above~; therefore, there is
 no physical ground to abstain from such an extension of the fitting range. 
  
\subsubsection{Testing The $\tau$ Spectrum Lineshapes}
\label{taulineshape}
 \indent \indent  The remarks presented in Subsection \ref{tauscale1} explain why
it is certainly
 appropriate to test the various $\tau$ spectrum lineshapes independently from their
 absolute magnitudes. This can be done in two different ways. 
 
 A first method turns out to normalize the data points $m_i$ to the sum
 of the data points covered by our fitted energy range (from threshold up to 1 GeV/c).
 Then, correspondingly, the model function function $f(s)$ on the left--hand side of Eq. (\ref{FF4})
 should be normalized to its integral over the fitted range.
 
 Another method, is simply to minimize the $\chi^2$  as defined by Eq. (\ref{FF6}),
 but amputated this from the $(\lambda/\eta)^2$ term which constrains the
 scale in accordance with the claimed experimental uncertainty. Indeed, in this
 way, the scale factor is allowed to vary freely within the global fit procedure.
 We checked that these two methods give similar results.
   
\subsubsection{Dealing With The Uncertainties In $e^+ e^-$ Data Samples}
\label{eeUncertainties}
\indent \indent Uncertainties in the $e^+ e^-$
data sets  are accounted for in several ways.  For the $e^+ e^- \ra \pi^0 \gamma$
data sets \cite{CMD2Pg2005,sndPg2000,sndPg2003} and the 
$e^+ e^- \ra \eta \gamma$ data sets 
\cite{CMD2Pg1999,CMD2Pg2001,CMD2Pg2005,sndPg2000,sndPg2007}, taking into
account the magnitude of the systematics, we did not find motivated  to split them up
into their bin--to--bin correlated and uncorrelated parts. We  just 
add in quadrature the reported systematic  and statistical errors. 

For the
$e^+ e^- \ra \pi^+ \pi^- \pi^0 $ samples, we dealt differently
with  the different data sets.  For the relatively unprecise
data sets   \cite{ND3pion-1991,CMD3pion-1989} we did as just explained
for $e^+ e^- \ra (\pi^0/\eta) \gamma$ data and simply added
systematic  and statistical errors in quadrature. 
Instead, for the more accurate data sets \cite{CMD2-1995corr,CMD2-2006,CMD2KKb-1,CMD2-1998}, 
only the uncorrelated part of the systematic uncertainty has been added in quadrature
to the statistical errors. On the other hand, the bin--to--bin correlated error
has been treated as emphasized above for the $\tau$ data sets. 
For reasons already emphasized in \cite{ExtMod1}, we have discarded
the $ e^+ e^- \ra \pi^+ \pi^-\pi^0$ data provided by
 \cite{SND3pionHigh,SND3pionLow}.

Finally, the various Novosibirsk $e^+ e^- \ra \pi^+ \pi^-$ data sets recently
collected \cite{CMD2-1995corr,CMD2-1998-1,CMD2-1998-2,SND-1998} carry a common
bin--to--bin {\it and} sample--to--sample correlated uncertainty of 0.4 \%.  The older
data from OLYA and CMD \cite{Barkov} also share a common correlated (scale)  uncertainty 
\cite{SimonPriv}
of $\simeq 1 $\%. In both cases, we have added the uncorrelated part of the
 systematics  to the statistical errors in quadrature.  
 
 Instead, in order to treat properly the correlated  uncertainty, one should  
 consider the data sets in   \cite{CMD2-1995corr,CMD2-1998-1,CMD2-1998-2,SND-1998}
 as a single (merged) data set and use as $\chi^2$ an expression like Eq. (\ref{FF6})
 to introduce the common scale to be fitted. Here also, if $\lambda/0.4\% \leq 1 \div 2$,
 one could remove this scale while performing the change  $V_{ij} \Rightarrow V_{ij} + (0.4\%)^2 m_i m_j $.
  One has   performed the same way, { \it mutatis mutandis}, with the older 
  OLYA and CMD data sets \cite{Barkov}.
  
 Because of the poor probability of the best fit to the KLOE data \cite{KLOE_ISR1} already 
commented upon in \cite{ExtMod1},
the corresponding  data sample is not included systematically in the 
set of  $e^+e^- \ra \pi^+ \pi^-$ data samples considered~; however, its  effects will be noted 
when relevant.
In order to fit the parameters of the IB $\rho,~\omg,~\phi$ mixing scheme \cite{taupaper,ExtMod1}, 
one still uses a subset
of  9  radiative decay width data which are taken  from the latest issue of the
Review of Particle Properties \cite{RPP2008} and are given explicitly in \cite{ExtMod1}.

\section{Simultanous Fits To $e^+ e^-$ and $\tau$ Data}
\label{FitProcess}
\subsection{Interplay Between $\delta m^2$, $\delta g$ And $\lambda$}
\label{preliminaire}
\indent \indent  Strictly speaking, the lineshape of the $\tau$ spectrum
is determined by the HLS parameters $g^\prime=g+ \delta g $ and the 
Higgs--Kibble mass 
$m^2 + \delta m^2$. This is clear from the expressions given in Section \ref{FitTau} above.
The specific Isospin breaking parameters $\delta m^2$ and $\delta g$  
differentiate the $\rho^\pm$ lineshape from those of the $\rho^0$ meson. 
However, these parameters also govern the absolute scale of the $\rho^\pm$ 
spectrum compared to the $\rho^0$ one. 
Therefore, if an uncertainty on the absolute scale of a 
 measured $\tau$ spectrum calls for a fit parameter
 $\lambda$  rescaling the whole data spectrum, it is quite important to
 examine its interplay with $\delta g$ and $\delta m^2$. 

\begin{table}[!htb]
\hspace{-1.0cm}
\begin{tabular}{|| c  ||  c   |c  ||  c | c  ||}
\hline
\hline
Data Set  & $\delta m^2$ (GeV$^2$) & $\delta g $ &  $\lambda$  & $[\chi^2/points]_{Exp}$ \\
\hline
 ALEPH \hhhv  &  $(3.37 \pm 1.27) ~10^{-3}$ 
&  $(-0.56 \pm 0.12)~10^{-1} $ 
 & $(-1.01 \pm 0.40) \%$ &  $ 27.16/38$    \\
\hline
BELLE \hhhv  &  $(-0.01 \pm 0.77) ~10^{-3}$ 
&  $(-0.12 \pm 0.10)~10^{-1} $ 
 & $(-3.83 \pm 0.54) \%$ &  $ 32.46/20$    \\
\hline
CLEO \hhhv  &  $(-1.53 \pm 1.07) ~10^{-3}$ 
&  $(0.16 \pm 0.14)~10^{-1} $ 
 & $(-5.51 \pm 0.74) \%$ &  $ 38.99/30$    \\
\hline
\hline
 ALEPH \hhhv  &  $(4.04 \pm 1.22) ~10^{-3}$ 
&  $(-0.69 \pm 0.11)~10^{-1} $ 
 & ${\bf 0}$ &  $ 29.19/37$   \\
\hline
BELLE \hhhv  &  $ (2.18 \pm 0.71) ~10^{-3}$ 
&  $(-0.51 \pm 0.08)~10^{-1} $ 
 & ${\bf 0} $ &  $ 41.12/19$    \\
\hline
CLEO \hhhv  &  $(2.26\pm 0.94) ~10^{-3}$ 
&  $(-0.54 \pm 0.11)~10^{-1} $ 
 & ${\bf 0} $ &  $ 61.49/29$    \\
\hline
\hline
\end{tabular}
\caption{
\label{T0} Global fit results with each $\tau$ data sample separately.
Bolface  numbers are actually not allowed to vary in the fits. Global fit
probabilities are always above 90\%. 
}

 \end{table}

For the present exercise, 
we consider all data sets 
involving  $e^+ e^- \ra \pi^+ \pi^-$ data together with all 
data sets covering the $ e^+ e^- \ra \pi^0 \gamma$ and $\eta \gamma$ 
annihilation channels. These additional channels
allow to remove the $\rho^0/\omg/\phi \ra (\pi^0/\eta) \gamma$ partial widths
from the vector meson decay mode subsample unavoidably used. In this Section, 
the ISR KLOE data sample for $e^+ e^- \ra \pi^+ \pi^-$ is removed
from the collection of data sets to be fitted~; we also leave aside
the $e^+ e^- \ra \pi^+ \pi^- \pi^0$ annihilation data which play a minor role
in defining the vector meson mixing scheme.

The  $e^+ e^-$ measurements with $s < 1.05$ GeV$^2$ are submitted 
to fit -- in order to include the $\phi$ region -- together
with all $\tau$ decay measurements  with $m_{\pi \pi} < 1$ GeV. 

We have performed simultaneous fits of each of the A, B and C $\tau$ data sets
together with the $e^+ e^-$ data referred to above.  The results are shown 
in Table \ref{T0} and exhibit quite interesting features, depending on the 
particular $\tau$ data set considered.

The first three lines provide parameter values when  $\delta m^2$, $\delta g$ and $\lambda$
are allowed to vary.  For each $\tau$ data sample,  $\lambda$ is constrained
by the relevant $[\lambda_{fit}/\eta_{Exp}]^2$ term in the global $\chi^2$.
In this case, one notes that~:
\begin{itemize}
\item The significance for a non--zero $\delta m^2$ is at $ \simeq 2.6 ~\sigma$ 
for A and negligible for B or C ,
\item The significance for a non--zero $\delta g$ is at the $ \simeq 1 ~\sigma$ 
level for for B or C but large for A ($4.7 \sigma$),
\item  The values for some important correlation  coefficients returned by the 
fit are large for each $\tau$ data set~:   
$(\delta g,\lambda)\simeq (\delta g,\delta m^2)\simeq - 50$ \% and 
$(\lambda,\delta m^2)\simeq (25 \div 50) $ \%. 
These values reflect the interplay between $\delta g$, $\delta m^2$ and $\lambda$
in determining the absolute scales of the experimental spectra.
\item  The significance for  non--zero $\lambda$'s is data set dependent~:
$2.5 ~\sigma_\lambda$  for A,
$7.1 \sigma_\lambda$ for B and $7.40 \sigma_\lambda$ for C. 
Compared with the scale uncertainties induced by the errors on the respective 
${\cal B}_{\pi \pi}$, this corresponds to $[\lambda_{fit}/\eta]_{ALEPH}=2.0 \pm 0.8$, 
$[\lambda_{fit}/\eta]_{BELLE}=2.5 \pm 0.35$ and $[\lambda_{fit}/\eta]_{CLEO}=3.2 \pm 0.43$.
Taking into account the large correlations, between $\delta g$, $\delta m^2$ and $\lambda$,
this looks to us acceptable.
\end{itemize}

\begin{figure}[!ht]
\begin{minipage}{\textwidth}
\begin{center}
\resizebox{\textwidth}{!}
{\includegraphics*{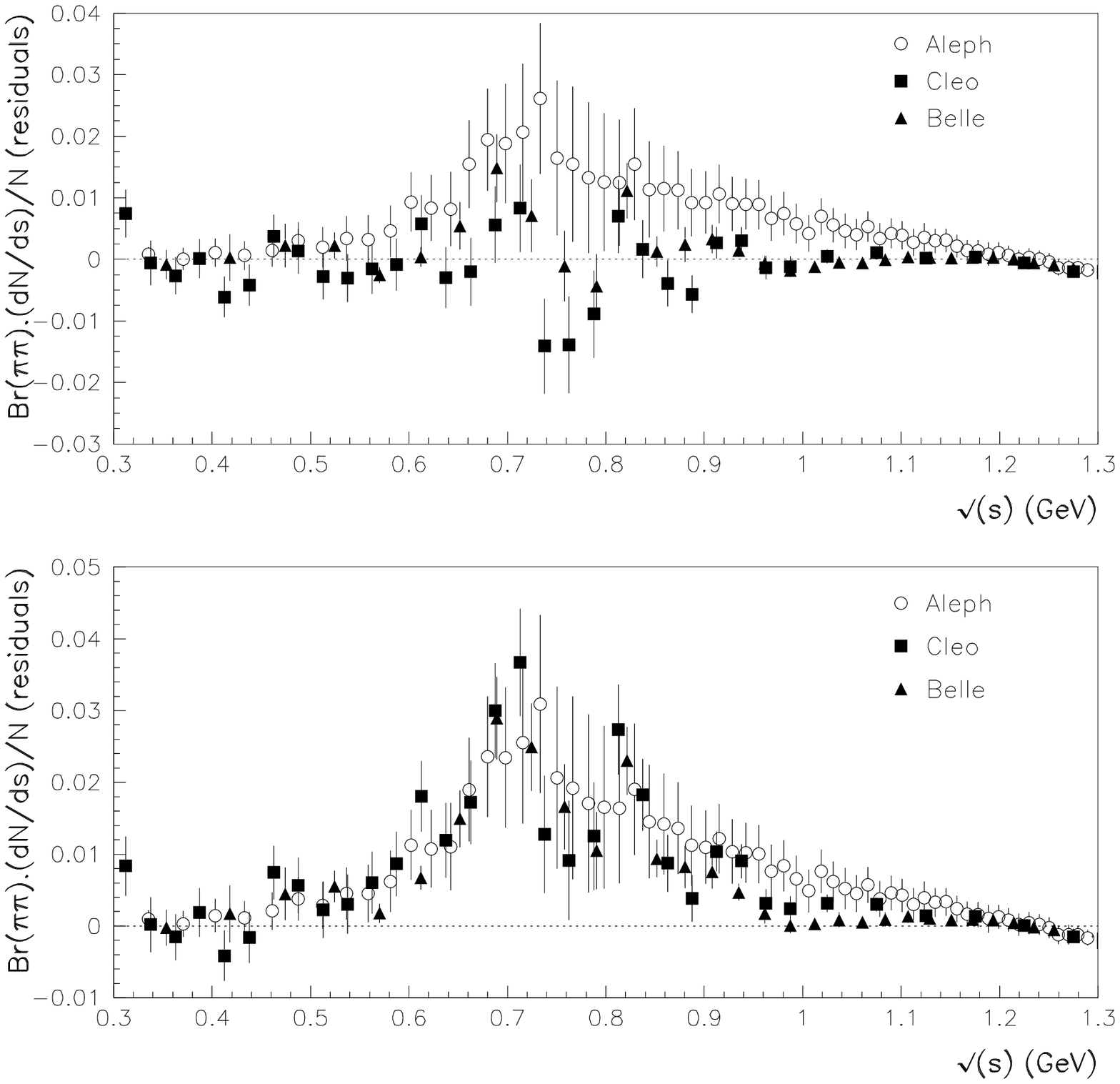}}
\end{center}
\end{minipage}
\begin{center}
\vspace{-0.3cm}
\caption{\label{dmdg} Residual distributions of ALEPH \cite{Aleph}, BELLE \cite{Belle} and 
CLEO \cite{Cleo} data sets in fits with mass and width differences between the charged and
neutral $\rho$ mesons. Upmost Figure includes fit rescaling factors contrained by the experimental
uncertainty on ${\cal B}_{\pi \pi}$~; downmost Figure corresponds to assuming no rescaling.
The fit region is bounded by 1 GeV/c.}
\end{center}
\end{figure}

The corresponding fit residuals are shown superimposed in the upmost Figure \ref{dmdg}. 
One clearly sees that the B and C  residuals are well spread around zero.  Those for A
are  slightly distorded around the $\rho$ peak in a way opposite to B and C. 
The last data column in Table \ref{T0} illustrates  that each of A, B and C
is well described by the global fit, simultaneously with $e^+ e^-$ data
(The so--called New Timelike data \cite{CMD2-1995corr,CMD2-1998-1,SND-1998}
always yield $\chi^2/points \simeq 118/127$).

\vspace{0.5cm}

The non--zero values for the $\lambda$'s could possibly mean that the reported 
measured values for ${\cal B}_{\pi \pi}$ are overestimated. However, the large 
correlations noted just above    prevent such strong
a conclusion, as the rescaling  effect could well be absorbed by
some different values for $(\delta g,\delta m^2) $.

It is, therefore, worth examining
the case when $\lambda \equiv 0$ is imposed. Of course, this turns
out to prevent any rescaling of the experimental spectra and, thus, check wether  the
mass and coupling breaking we introduced could
account alone for the measured reported normalizations of the various $\tau$ spectra. 
The corresponding results are shown 
in the downmost three lines of Table \ref{T0} and call for important remarks~:
 
\begin{itemize}
\item  The significance for a non--zero mass shift $\delta m^2$ is  always
improved~: $3.6~\sigma$ for A, $3.1~\sigma$ for B and
$2.4~\sigma$ for C,
\item  The coupling difference always becomes highly significant~: 
the ratio of the central value to its uncertainty is 6.3 for A and
B and 4.9 for CLEO.
\item  However, fixing $\lambda \equiv 0$ marginally degrades the
BELLE description, more significantly the account of
 CLEO data, whereas the fit quality is unchanged -- and quite good -- 
 for ALEPH data.
\end{itemize}
The description of $e^+ e^-$ data are, in this case, marginally degraded (The
new timelike data, for instance yield $\chi^2 \simeq 125$ instead of 118 before). 

\vspace{0.5cm}

This leads us to conclude that the additional Isospin Symmetry breaking 
mechanism, which summarizes into   mass and width differences for the $\rho$'s,
allows to account for the original absolute normalization of the ALEPH spectrum with
a very good probability. One can thus conclude that the dipion ALEPH spectrum
-- or, equivalently, the ALEPH pion form factor --  is in in full accord with VMD predictions,
{\it
provided one suitably accounts for   mass and width shifts in the $\rho^\pm$
information  compared to $\rho^0$}.
Numerically, this turns out to plug into  the ALEPH spectrum parametrization~:
\be
\left\{
\begin{array}{llll}
\displaystyle \frac{\Gamma_{\rho^\pm}-\Gamma_{\rho^0}}{\Gamma_{\rho^0}} \simeq &
\displaystyle \frac{2 \delta g}{ g} =& [-2.50 \pm 0.40] ~\%\\[0.5cm]
\displaystyle m^2_{\rho^\pm}-m^2_{\rho^0}  =& \delta m^2=&(4.04 \pm 1.22) ~10^{-3} 
~~{\rm GeV}^2
\end{array}
\right .
\label{rhobrk}
\ee
which approximately\footnote{Mass and width of broad objects like
the $\rho$ are, conceptually, 
definition dependent  and, additionally, we have not accounted for the large
negative correlation term  $(\delta g,\delta m^2)$.
Moreover, these numerical values somewhat vary, depending on the exact $e^+ e^-$
data set content submitted to the global fit.} 
means $m_{\rho^\pm}-m_{\rho^0} \simeq 2.59 \pm 0.78$ MeV
and $\Gamma_{\rho^\pm}-\Gamma_{\rho^0} \simeq -3.7 \pm 0.6 $ MeV. If $\delta m^2$,
 is numerically in the usual ballpark \cite{RPP2008},
$\delta g$  seems slightly larger than could have been inferred 
from commonly reported estimates \cite{RPP2008,Ghozzi}
for $\Gamma_{\rho^\pm}-\Gamma_{\rho^0}$.

The results corresponding to fits with $\lambda=0$ are shown in the downmost Figure
\ref{dmdg}. Compared with the upmost one, this Figure shows interesting features~:
the ALEPH residual distribution is essentially unchanged, but the residuals for BELLE and 
CLEO are shifted towards positive values and start resembling the ALEPH residual
distribution. This indicates
graphically that some rescaling is still needed in order to
get a satisfactory description of the B and C data sets. Whether this residual rescaling
could be absorbed into a more sophisticated Isospin breaking distortion mechanism
cannot be discarded.

As a summary, one can assert that, with appropriately chosen $\delta m^2$ and $\delta g$, 
ALEPH data can avoid any significant rescaling within the model we developped. 
This attractive property is
unfortunately not completely shared by BELLE and CLEO in isolation. This is reflected 
by the fit information given in the last data column of Table \ref{T0} and
by the downmost Figure \ref{dmdg}.  Nevertheless, with some rescaling, the BELLE and CLEO
data sets can be satisfactorily understood.

Moreover, it is clear from the above analysis that there is a large interplay
between parameters defining IB shape distortions and the absolute scale
of the experimental spectra.  Stated otherwise, some part of a fitted 
rescaling could be due to the difficulty to model the actual shape distortions 
and, conversely, some part of a real scale factor could well be absorbed by a
fitted distortion. Therefore, one cannot claim, in view of a large fitted
rescaling, that the reported ${\cal B}_{\pi \pi}$ should be rescaled by as much.

\subsection{Fitting The $\tau$ Lineshapes}
\indent \indent
In view of the different behaviors of the A, B and C data sets, it is worth
examining separately the  $\tau$ spectrum lineshapes besides the full spectra. 
As explained in Subsection \ref{taulineshape}, this turns out to fit the
$\tau$ data samples without including the scale constraining terms in their
contributions to the total $\chi^2$.

For this purpose, one has first performed a global fit leaving free $\delta m^2$,
$\delta g$ and the $\lambda$'s unconstrained. One thus gets 
$\delta g=(0.15 \pm 0.08) ~10^{-1}$
and $\delta m^2 = (-1.47 \pm 0.67) ~10^{-3}$ GeV$^2$ 
with\footnote{The number of data points
in the $\tau$  data samples submitted to fit are always 37 (A), 19 (B) and 29 (C).}
 $\chi^2_{ALEPH}=17.86$, $\chi^2_{BELLE}=29.37$ and $\chi^2_{CLEO}=29.71$ associated
 with  a $\chi^2=118.31$ for the 127 data points of the so--called New Timelike data
\cite{CMD2-1995corr,CMD2-1998-1,CMD2-1998-2,SND-1998}. Instead, when imposing
$\delta g=\delta m^2 =0$, one gets 
 $\chi^2_{ALEPH}=19.22$, $\chi^2_{BELLE}=27.61$ and $\chi^2_{CLEO}=31.83$
 while the New Timelike data get  $\chi^2=120.23$. 
 
 One may compare
 the $\chi^2$ values yielded for each of the $\tau$ data samples
 when fitting together their normalized spectra with those already 
 reported\footnote{Let us remind that the fit results reported in
 this Table are, instead, obtained from (global) fits  using
 each of the $\tau$ data sets in isolation. } in Table \ref{T0}.
 
 The results given just above enforce the conclusions we reached in
 Subsection \ref{preliminaire}. In the global fit procedure, 
 it looks difficult to distinguish effects due to IB shape distortions
 in the $\rho$ distribution from genuine rescaling effects. 
 This is especially striking when considering the ALEPH
 data set~: One gets a quite acceptable description of the invariant mass
 distribution by either assuming $\lambda=0$ and $(\delta g,~\delta m^2)$
 free or by letting $\lambda$ free and imposing $\delta g=\delta m^2=0$. So, 
 the reported scale discrepancy can be absorbed~; however, the
 sharing of the effect between true (physical) lineshape distortions
 and an (experimental) bias in the accepted value for ${\cal B}_{\pi \pi}$
 is beyond the scope of  fit results. This is
 an important issue for using $\tau$ data in order to estimate
 hadronic contributions to $g-2$~; this will be commented on at the appropriate
 place.

Therefore, at least when fitting the $\tau$ lineshapes, it is not justified to keep
non--vanishing $\delta m^2$ and $\delta g$. In this case, the main difference between the 
$\rho^+$ and the (physical) $\rho^0$ is essentially carried  \cite{taupaper,ExtMod1}  by  
the $\gamma- \rho^0$ and the $W-\rho^\pm$ transition amplitudes 
$f_\rho^\gamma$ and $f_\rho^\tau$~:
\be
\frac{f_\rho^\gamma}{f_\rho^\tau}=1 + \frac{1}{3} \alpha(s)
+\frac{z_V \sqrt{2}}{3} \beta(s) ~~~~,~~(f_\rho^\tau=ag^2f_\pi^2)
\label{referee1}
\ee

These amplitudes differ 
through terms explicitly depending on the Isospin Symmetry breaking functions which account for the vector
meson mixing \cite{taupaper,ExtMod1}. The $s$--dependent terms in Eq. (\ref{referee1})
account for the isospin zero component of the $\rho^0$ meson. 
Keeping non--vanishing $\delta g$ and $\delta m^2$ within our modelling, 
the right--hand side of Eq. (\ref{referee1}) gets a leading 
additional $(\delta m^2/m^2 - \delta g/g)$ term. 

\begin{table}[!htb]
\hspace{-2.4cm}
\begin{tabular}{|| c  || c || c   |c  ||  c | c  ||}
\hline
\hline
\hhhu ~~ & Expected & \multicolumn{2}{|c|}{ No $\tau$ Data}  &  \multicolumn{2}{|c|}{$\tau$ Data Set Configurations} \\
\hhhu & r.m.s. & $e^+e^-$ NSK+KLOE & $e^+e^-$ NSK &  NSK + A$^{sh}$B$^{sh}$C$^{sh}$ & NSK + ABC \\
\hline
\hline
Scale New Timelike \hhhv& $0.4\%$  &  $(-0.9 \pm 0.4) \%$      &  $(-0.6 \pm 0.4) \%$  
& $(-0.6 \pm 0.4)\%$  &
  $ (+0.4 \pm 0.3)$ \% \\
\hline
Scale Old Timelike \hhhv     & $1.0\%$ &   $(+1.4 \pm 0.7) \% $  &  $(+1.6 \pm 0.7) \% $   
& $(+1.6 \pm 0.7)\%$ &  
$ (+2.4 \pm 0.7)$  \%\\
\hline
Scale CMD--2 \cite{CMD2-1995corr} \hhhv  &  $ 1.3 \%$ & $(-0.4 \pm 1.2)\%$ & $(-0.5 \pm1.2)\%$ 
  & $(-0.4 \pm  1.2)
\%$ &  $ (-0.7 \pm 1.2)$  \%\\
\hline
Scale CMD--2 \cite{CMD2-2006} \hhhv & $2.5\%$  &  $(-2.6 \pm 1.9)\%$   
&  $(-1.8 \pm 1.9)\%$   & $(-1.8 \pm 1.9)\% $ &  $
(-1.4 \pm 1.9)$  \%  \\
\hline
Scale CMD--2 \cite{CMD2KKb-1} \hhhv & $4.6\%$  &  $(-4.7 \pm 3.4) \%$ 
&  $(-4.2 \pm 3.4) \%$ 
 & $(-4.1 \pm 3.5) \%$ &  $ (-3.6 \pm 3.4)$  \%  \\
\hline
Scale CMD--2 \cite{CMD2-1998} \hhhv & $1.9\%$  &  $(-2.6 \pm 1.6) \%$  & $(-2.2 \pm 1.6) \%$  
& $(-2.0 \pm 1.6)\%$ &
 $ (-2.1 \pm 1.6)$  \%  \\
\hline
$g$ \hhhv &---  & $5.569 \pm 0.003$  &   $5.587 \pm 0.008$   & $5.565 \pm 0.006$ 
&  $5.532 \pm 0.007$ \\
\hline
$\delta g$ \hhhv &---  &  --- &   ---   & ${\bf 0}$ &  
$[-0.30 \pm 0.07]~10^{-1}$ \\
\hline
$\delta m^2$ (GeV$^2$)\hhhv &---  &  --- &   ---   & ${\bf 0}$  
&  $[1.09 \pm 0.6]~10^{-3} $ \\
\hline
$x$ \hhhv &---  & $0.917 \pm 0.012$  &   $0.908 \pm 0.013$   & $0.905 \pm 0.013$ 
&  $0.902 \pm0.013$ \\
\hline
$z_A$ \hhhv &---  &  $1.501 \pm 0.010$   &   $1.476 \pm 0.009$   
& $1.468 \pm 0.009$ &  $1.454 \pm 0.009$ \\
\hline
$a$ \hhhv &---  & $2.372 \pm 0.002$  &   $2.357 \pm 0.005$   & $2.361 \pm 0.004$ 
&  $2.382 \pm 0.004$ \\
\hline
$c_3$ \hhhv &--- & $0.927 \pm 0.006$ &   $0.936 \pm 0.006$   & $0.944 \pm 0.006$ 
&  $0.952 \pm 0.006$ \\
\hline
$c_1-c_2$ \hhhv &---  &$1.194 \pm 0.031$ &    $1.196 \pm 0.032$   & $1.228 \pm 0.032$ & 
$1.237 \pm 0.032$ \\
\hline
\hline
$\chi^2/\rm{dof}$ \hhhu &  --- &  647.43/653&  521.15/593 &602.36/675 
&645.22/676 \\
Probability  & ---  & 55.4\% &98.5\%  & 97.9\% & 79.7\% \\
\hline
\hline
\end{tabular}
\caption{
\label{T1} Results in various fit configurations. $e^+e^-$ NSK data stand for all annihilation processes 
discussed in the text. Inclusion of ALEPH, BELLE and CLEO data are referred to as A, B and C
spectra respectively, A$^{sh}$, B$^{sh}$ and C$^{sh}$ denote the corresponding
normalized spectra. The upper part
displays the rescaling factors for $\pi^+ \pi^-$ (first two lines) and $\pi^+ \pi^- \pi^0$ data 
samples. The lower part refers to breaking parameters discussed in Ref. \cite{ExtMod1}.
Numbers written boldface are parameter values not allowed to vary inside fits.
}

 \end{table}

\vspace{1.cm}

From now on, we choose to reintroduce the $e^+ e^- \ra \pi^+ \pi^-\pi^0$
data from  \cite{ND3pion-1991,CMD3pion-1989}  and 
\cite{CMD2-1995corr,CMD2-2006,CMD2KKb-1,CMD2-1998}, as
discussed in Subsection \ref{eeUncertainties} above.
We also let the scale factors vary for both the $e^+ e^-$ and $\tau$ data
sets\footnote{keeping, of course, the term constraining the scale variation
for all the relevant data sets.}. Some numerical  results
obtained with various data set configurations  are reported in Table \ref{T1}.  
One notes that all scale factors introduced for the $e^+ e^- \ra \pi^+ \pi^-$
and $e^+ e^- \ra \pi^0 \pi^+ \pi^-$ cross sections are in good correspondence with 
the expectations reminded in the first data column. This result is understood as  
confirming once more \cite{ExtMod1} the claims of the corresponding experiments
concerning their correlated systematic uncertainties. 
In the same Table, the data column A$^{sh}$B$^{sh}$C$^{sh}$ reports the results obtained while 
removing the scale constraining terms (the second one in Eq. (\ref{FF6}) for each $\tau$ data sample).
The numerical values of the rescaling parameters for the $\tau$ spectra are given below.

Table \ref{T1} justifies the removal of the  $e^+ e^-$ rescaling 
factors\footnote{Then, the correlated scale uncertainties only appear in the 
full error covariance matrix as reminded in Subsection \ref{eeUncertainties} and in Section 6 of 
\cite{ExtMod1}.} from the set of parameters to be fitted.
One also observes that the physics parameters have only (small) reasonable fluctuations.
Finally,  Table \ref{T1} clearly shows that all Novosibirsk (NSK) data (4 different 
cross sections) are quite consistent with the freely rescaled $\tau$ decay data as
the fit yields a probability above the 90\% level.
Moreover, it is also interesting to note
that constraining the rescaling of the $\tau$ spectra by their reported scale uncertainty 
({\it i.e.}, keeping the scale fixing terms as in Eq. (\ref{FF6})) still
provides quite comfortable probabilities (above the 80\% level), as can be seen from
the last data column in Table \ref{T1}.

This last data column exhibits also interesting features~: In a simultaneous fit
to all $\tau$ data samples together with $e^+ e^-$ data and vector meson 
decay modes, the significance for  a non--vanishing
$\delta g$ is $\simeq 4 \sigma$ whereas that for $\delta m^2$ is only
 $\simeq 1.8 \sigma$. Comparing with the  separate  $\tau$ data sample
 fits reported in Table \ref{T0}  indicates that significantly
 non--vanishing $\delta m^2$ and $\delta g$ are driven by ALEPH data only. 

We have explored the effect of including
the KLOE data set \cite{KLOE_ISR1} beside the whole set of Novosibirsk $e^+ e^-$ data
and the three $\tau$ decay data sets. As expected from its intrinsically large $\chi^2$
 \cite{ExtMod1} (123  for 60 measurements in the present case), 
the global fit probability drops to $\simeq 8$\%.
However,  the information displayed in Table
 \ref{T1} is not substantially modified. In particular, the scale factors
 affecting the $e^+ e^-$ Novosibirsk data remain very close to the expectations
 reported in the first data column.

From now on, all scale factors affecting the $e^+e^-$  data 
-- except for KLOE, when relevant -- are set to zero and
the scale uncertainties are transfered into the error covariance matrix as
explained above and emphasized in Section 6 of \cite{ExtMod1}.

\subsection{Final Global Fits To $e^+ e^-$ and $\tau$ Data}
\label{FinalFit}
\indent \indent In this Subsection, we only refer to fits performed without
the KLOE data set. Some results are displayed in Table \ref{T2} and will be discussed 
now~; other pieces of information will be given later on,  
especially those with each  $\tau$ data sample fitted in isolation
with the whole set of $e^+e^-$ data. 

Two kind of results are shown in Table \ref{T2}, the former with fitting only the
$\tau$ spectrum lineshapes\footnote{In this case, the $\tau$ data are flagged
as A$^{sh}$, B$^{sh}$ and C$^{sh}$ for respectively the ALEPH, BELLE and CLEO data sets.
When using, for each $\tau$ data set, the full Eq. (\ref{FF6}), these data
sets are simply flagged A, B and C. } (see Subsection \ref{taulineshape} above),
the later with the full  $\tau$ spectrum information as expressed by 
Eq. (\ref{FF4}).  The first data column in Table \ref{T2} displays the results
obtained with only the whole set of $e^+e^-$ data \cite{ExtMod1}.

\begin{figure}[!ht]
\begin{minipage}{\textwidth}
\begin{center}
\resizebox{\textwidth}{!}
{\includegraphics*{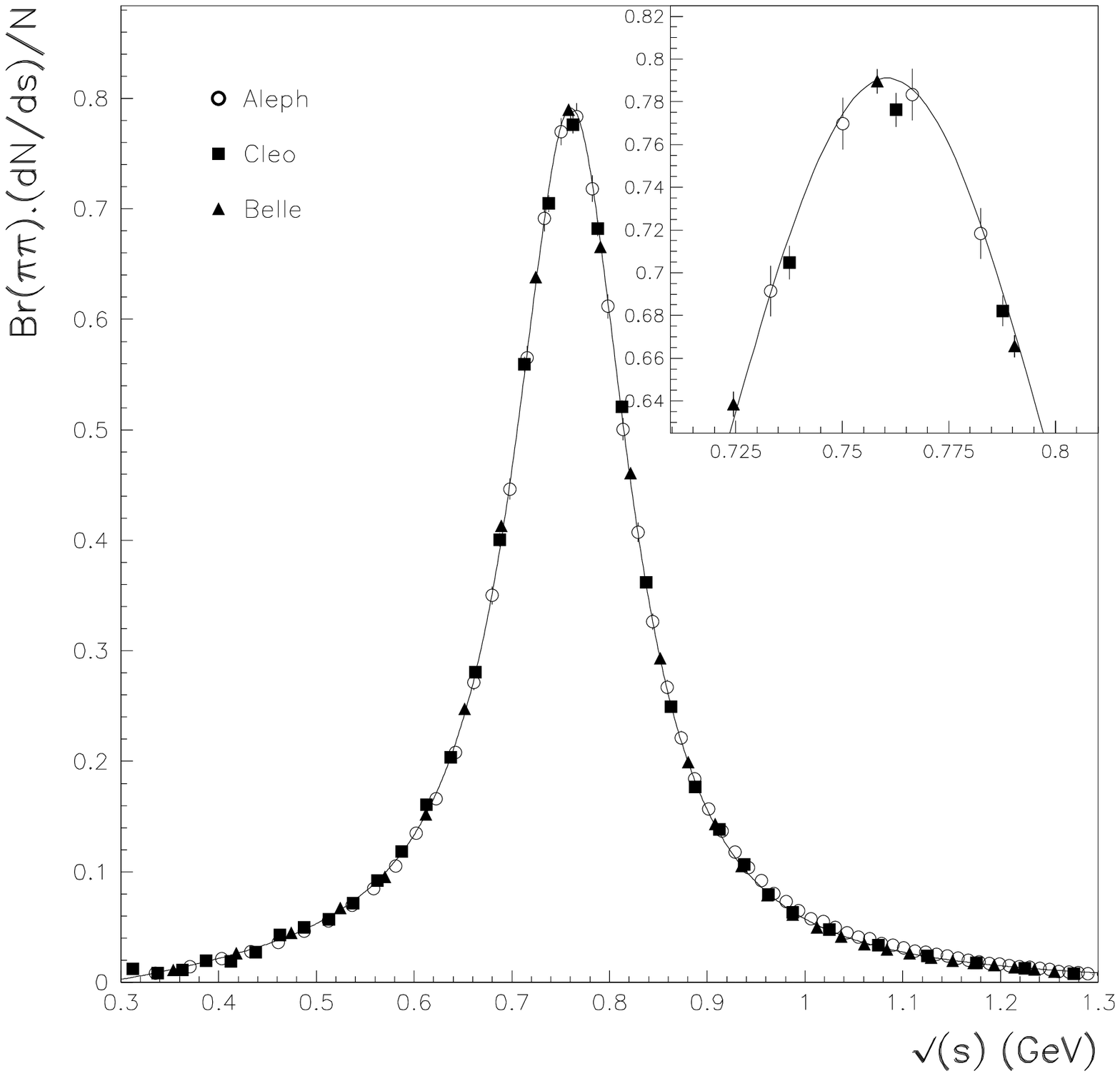}}
\end{center}
\end{minipage}
\begin{center}
\vspace{-0.3cm}
\caption{\label{Tau_norm} Fit to the $\tau$ spectra 
from ALEPH \cite{Aleph}, BELLE \cite{Belle} and CLEO \cite{Cleo} data sets. 
The absolute normalization  of the $\tau$ spectra is completely free.
}
\end{center}
\end{figure}

\begin{figure}[!ht]
\begin{minipage}{\textwidth}
\begin{center}
\resizebox{\textwidth}{!}
{\includegraphics*{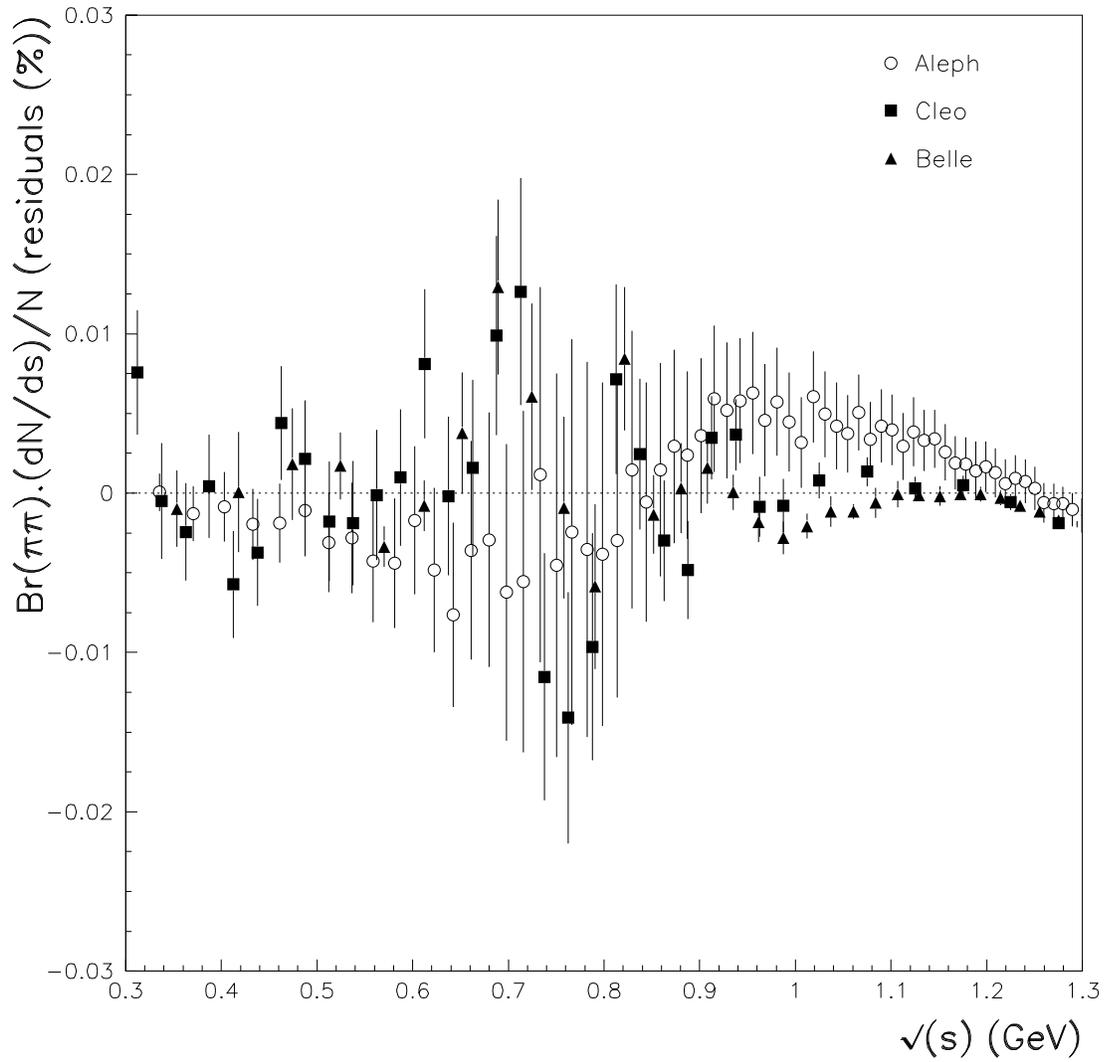}}
\end{center}
\end{minipage}
\begin{center}
\vspace{-0.3cm}
\caption{\label{ResTau_norm} Residuals distribution in the fit to the $\tau$ 
spectra from  ALEPH \cite{Aleph},
BELLE \cite{Belle} and CLEO \cite{Cleo} data sets. 
The absolute normalization  of the $\tau$ spectra is completely free.
}
\end{center}
\end{figure}

The first two data columns in Table \ref{T2} clearly illustrate that the fit quality 
remains optimum\footnote{The origin of such favorable probabilities is discussed at the end of
this Subsection.} (above the 90\% level) when including only the $\tau$ spectrum 
{\it lineshapes} inside the fit procedure (data column flagged with A$^{sh}$ B$^{sh}$ C$^{sh}$).
This does not hide some badly described data set as reflected by the various partial
$\chi^2$ displayed.  Figure (\ref{Tau_norm}) displays
the normalized spectra together with the fit function~; the residual distributions 
shown in Figure (\ref{ResTau_norm}) confirm the goodness of the fit.

Comparing the individual $\chi^2$ in the second data column with those in the
first one for the different $e^+e^-$ data subsets
(and the decay width data set), one clearly observes negligible modifications
of their fit quality.   Moreover, the $\chi^2$ obtained for the
individual A$^{sh}$, B$^{sh}$, C$^{sh}$  spectrum lineshapes are also quite reasonable.

The numbers shown  boldface and within parentheses in the first data column 
are also quite interesting. Fixing the arbitrary scales to their fit values
(given in second data column) for each $\tau$ data set, one can {\it compute}
the  $\chi^2$ distance of each of the A$^{sh}$, B$^{sh}$, C$^{sh}$  spectrum lineshapes to the
fit solution to {\it only} the  $e^+ e^-$ data. Comparing these numbers to
their homologues in the second column (obtained, instead, with the $\tau$ spectrum lineshapes 
fitted) clearly proves that the $e^+ e^-$ 
data allow to {\it predict} the $\tau$  spectrum lineshape with
very good accuracy\footnote{We already reached this 
conclusion \cite{taupaper} with the CLEO data sample lineshape. }. 
One should note that, if unconstrained,
all $\tau$ spectrum normalizations prefer consistently a rescaling by $\simeq -5 \%$.
As discussed in Subsection \ref{preliminaire} above, the exact meaning of such rescalings 
should be considered with great care.   

\begin{figure}[!ht]
\begin{minipage}{\textwidth}
\begin{center}
\resizebox{\textwidth}{!}
{\includegraphics*{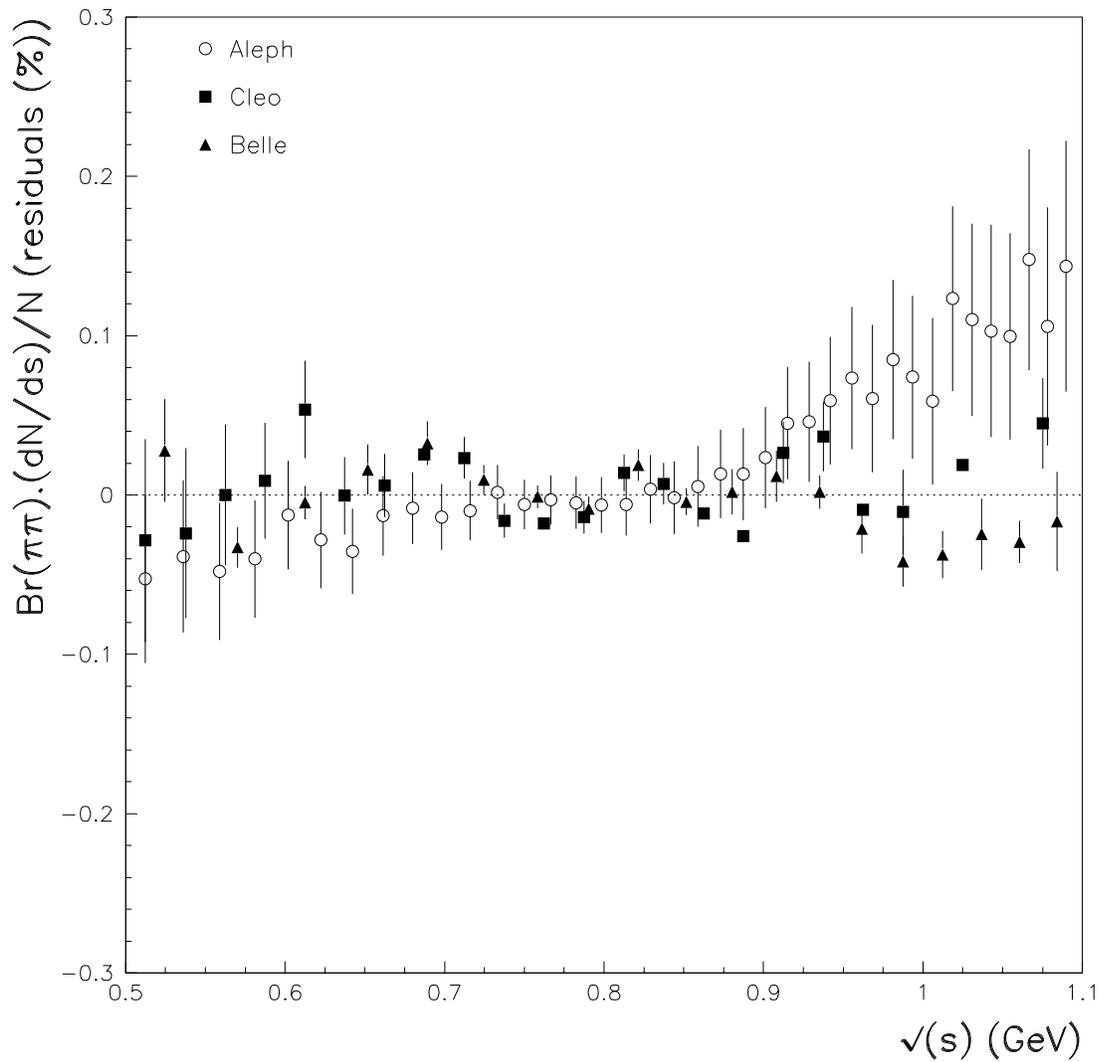}}
\end{center}
\end{minipage}
\begin{center}
\vspace{-0.3cm}
\caption{\label{ResTau_norm_spect} Residuals distribution in the fit to the $\tau$ 
spectra from  ALEPH \cite{Aleph}, BELLE \cite{Belle} and CLEO \cite{Cleo} data sets.
The residuals are normalized too the fit function values.
The absolute normalization of the $\tau$ spectra was left completely free.
}
\end{center}
\end{figure}

In Figure \ref{ResTau_norm_spect} we plot the fit residuals normalized to the fit 
function value, namely~:
\be
x_i=\displaystyle \frac{(1+\lambda) m_i - f_i}{f_i}~~~,~~ i=1,N_{Exp}
\label{FF8}
\ee
neglecting the effects of correlations on the plotted errors. This plot
clearly shows that, up to $\simeq 1$ GeV, the residuals can be considered as
 flat, structureless,  distributions~;  
Figure   \ref{ResTau_norm_spect}  is the exact analog of Figure 12 in \cite{Belle}
and exhibits the similar character of BELLE and CLEO data in our fitted 
$s$ range.

\begin{table}[!htb]
\hspace{+1.0cm}
\begin{tabular}{|| c  || c  || c  | c || }
\hline
\hline
\hhhu Data Set  & No $\tau$ Data &  \multicolumn{2}{|c|}{$\tau$ Data Set Configurations} \\
\hhhu ($\sharp$ data points) & NSK($e^+e^-$)   & NSK ${\cal +}$A$^{sh}$B$^{sh}$C$^{sh}$  & NSK  ${\cal +}$  ABC \\    
\hline
\hline
Decays (9) \hhhv       & $14.84$  & $15.12$  & $16.20$  \\
\hline
New  Timelike (127)\hhhv & $118.88$   & $120.24$  &  $126.47$ \\
\hline
Old Timelike (82)  \hhhv  & $50.65$ &  $51.08$   & $60.45$  \\
\hline
$\pi^0 \gamma$ (86)\hhhv & $66.03$  &  $65.84$  &  $66.07$ \\
\hline
$\eta \gamma$ (182)\hhhv  & $135.26$ &    $135.54$  &  $135.78$  \\
\hline
$\pi^+ \pi^- \pi^0$ (126) \hhhv  & $139.45$ &    $139.42$ &  $139.44$   \\
\hline
\hline
$\delta g$ \hhhv& --- & ${\bf 0}$& $[-0.30 \pm 0.07]~10^{-1}$ \\
\hline
$\delta m^2$ (GeV$^2$)\hhhv & --- & ${\bf 0}$ &$[1.06 \pm 0.62]~10^{-3}$ \\
\hline
\hline
ALEPH (37+1) \hhhv  &{ \bf (23.52)}&   19.23 & 28.30/36.51 \\
\hline
ALEPH Scale (\%) \hhhv  & --- & $-5.45 \pm 0.62$  &  $-1.46 \pm 0.38$ \\
\hline
\hline
CLEO (29+1) \hhhv  & { \bf (35.57)}  &    31.83 & 37.22/39.46  \\
\hline
CLEO Scale (\%) \hhhv  & --- & $-5.48 \pm 1.01$   & $-2.60\pm 0.54$  \\
\hline
\hline
BELLE (19+1) \hhhv  & { \bf (27.16)} &  27.61& 26.43/28.29\\
\hline
BELLE Scale (\%) \hhhv  & --- &  $-4.80 \pm 0.71$   & $-2.09 \pm 0.46$   \\
\hline
\hline
$\chi^2/\rm{dof}$ \hhhu & 525.10/597     & 605.90/679  & 648.68/680  \\
Probability  &   98.4\%  &  97.9\%   &  80.1\% \\
\hline
\hline
\end{tabular}
\caption{
\label{T2} Individual $\chi^2$ of the data samples under various fit
configurations. The $e^+ e^- $ data sets are referred
to in the text. Inclusion of ALEPH, BELLE and CLEO data sets are referred to 
in column flags as A, B and C, respectively. A$^{sh}$, B$^{sh}$, C$^{sh}$ correspond
to fitting the lineshape of the normalized invariant mass spectra only.
Numbers indicated boldface are the $\chi^2$ distances of A$^{sh}$, B$^{sh}$, C$^{sh}$
to the fit solution of {\it only} the $e^+ e^- $ data. Parameter values written boldface
are not allowed to vary in fits.
}
 \end{table}

\vspace{0.5cm}

The last data column in Table \ref{T2} displays the fit information while
fitting the ${\cal B}_{\pi\pi}/N~dN/ds$ distributions of the various 
$\tau$ experiments together with the whole set of $e^+e^-$ data, taking
into account the constraint imposed on the $\lambda$ scale factors by
the uncertainty on the measured branching ratio values ${\cal B}_{\pi\pi}$.

The full fit thus obtained is also quite satisfactory as reflected by the fit
probability (80\%). It shows, nevertheless,
that a significant rescaling of the experimental data is requested. For the 
$\tau$ data sets, the partial $\chi^2$ information is displayed as $x/y$ where 
$x$ is the part of the $\chi^2$ coming from the data points and $y-x$ is the contribution
of the $(\lambda/\eta)^2$ term (see Eq. (\ref{FF6})). $\sqrt{y-x}$ tells how far 
from the ${\cal B}_{\pi \pi}$  central value --
in units of $\eta_{Exp}$ --  , the spectrum normalization is preferred. 
Assuming no external cause to the rescaling, 
 the numerical values of the corresponding coefficients  look 
close to the expected ${\cal B}_{\pi \pi}$ value, taking into account the reported accuracy 
of the various measurements. Compared to these, the fit values are found at resp.
2.9 $\sigma_{ALEPH}$, 1.4 $\sigma_{BELLE}$ 
and  1.5 $\sigma_{CLEO}$ towards lower central values for ${\cal B}_{\pi \pi}$. 

\begin{figure}[!ht]
\begin{minipage}{\textwidth}
\begin{center}
\resizebox{\textwidth}{!}
{\includegraphics*{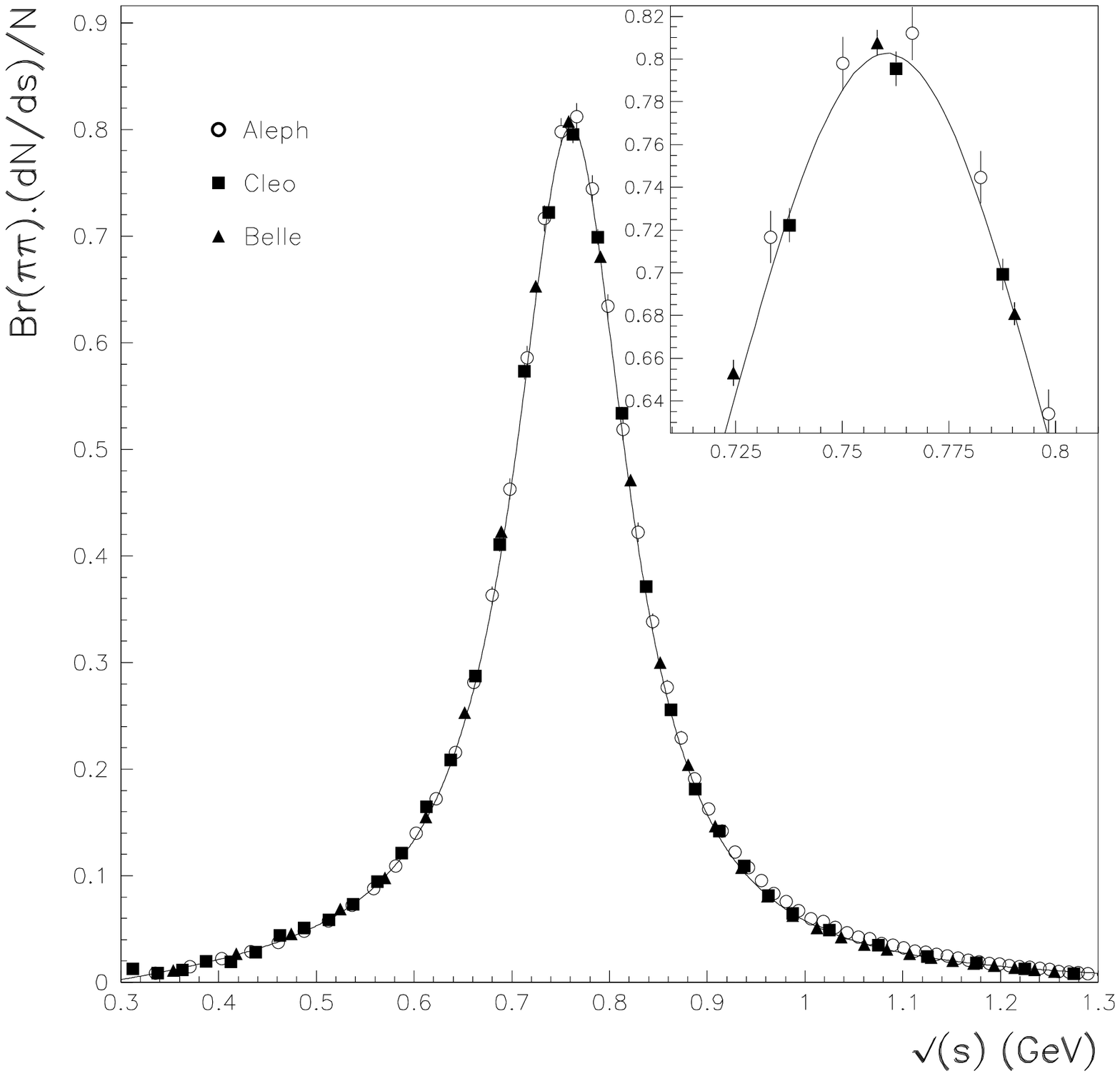}}
\end{center}
\end{minipage}
\begin{center}
\vspace{-0.3cm}
\caption{\label{Tau_abs} Fit to the $\tau$ spectra 
from ALEPH \cite{Aleph}, BELLE \cite{Belle} and CLEO \cite{Cleo} data sets. 
The normalization is fit but constrained by the experimental uncertainties
on ${\bf B}_{\pi\pi}$ (see text).
}
\end{center}
\end{figure}

\begin{figure}[!ht]
\begin{minipage}{\textwidth}
\begin{center}
\resizebox{\textwidth}{!}
{\includegraphics*{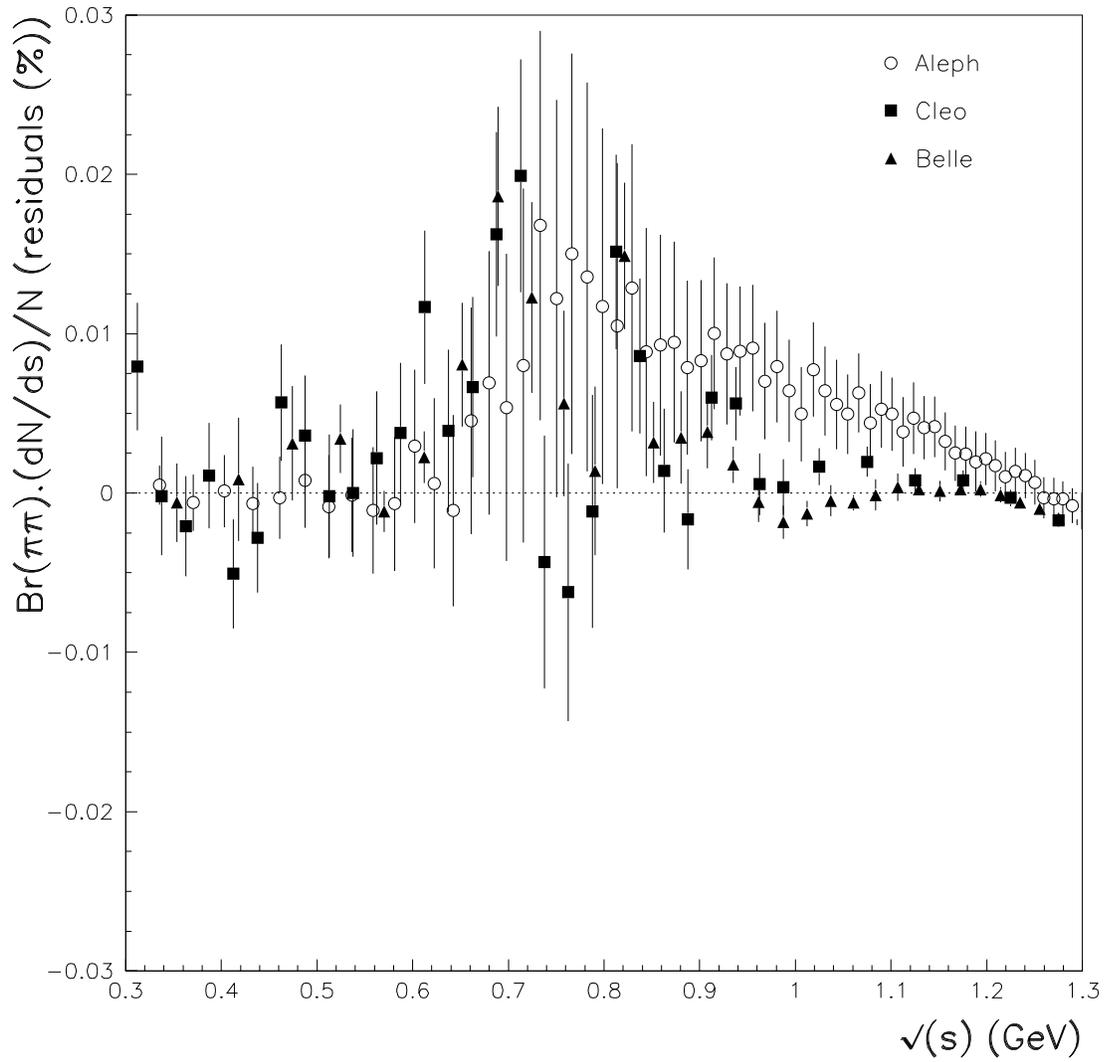}}
\end{center}
\end{minipage}
\begin{center}
\vspace{-0.3cm}
\caption{\label{ResTau_abs} Residuals distribution in the fit to the $\tau$ 
spectra from  ALEPH \cite{Aleph},
BELLE \cite{Belle} and CLEO \cite{Cleo} data sets. 
The normalization is fit but constrained by the experimental uncertainties
on ${\bf B}_{\pi\pi}$ (see text).
}
\end{center}
\end{figure}

\begin{figure}[!ht]
\begin{minipage}{\textwidth}
\begin{center}
\resizebox{\textwidth}{!}
{\includegraphics*{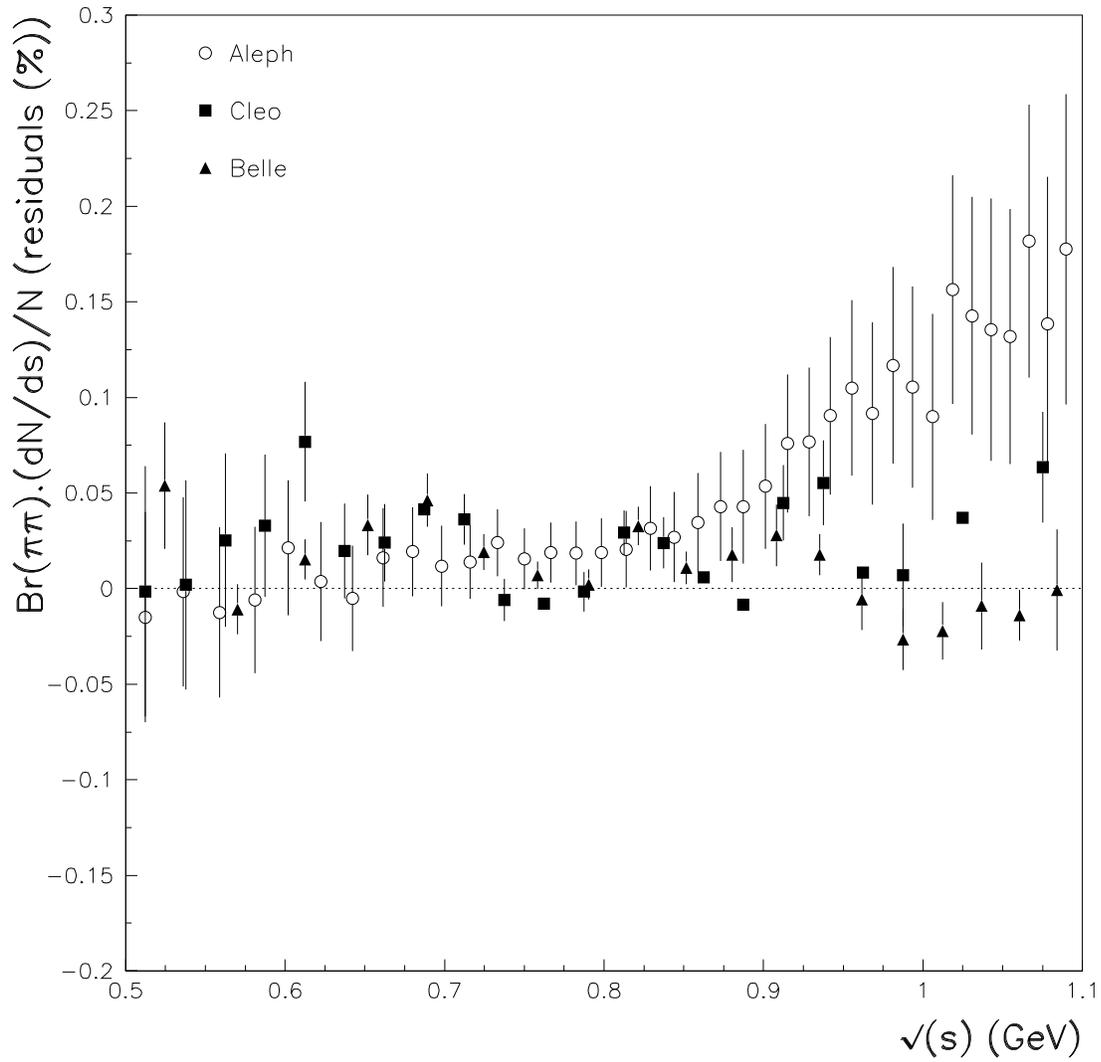}}
\end{center}
\end{minipage}
\begin{center}
\vspace{-0.3cm}
\caption{\label{ResTau_abs_spect} Residuals distribution in the fit to the $\tau$ 
spectra from  ALEPH \cite{Aleph}, BELLE \cite{Belle} and CLEO \cite{Cleo} data sets.
The residuals are normalized too the fit function values.
The absolute normalization of the $\tau$ spectra was constrained by the uncertainty on
${\bf B}_{\pi\pi}$ (see text).
}
\end{center}
\end{figure}

The results are represented in Figures (\ref{Tau_abs}),
(\ref{ResTau_abs}) and (\ref{ResTau_abs_spect}). In Figure (\ref{ResTau_abs})
one clearly sees that the residuals for CLEO and BELLE
are consistent with those in Figure (\ref{ResTau_norm}), while those for ALEPH
exhibit a structure around the $\rho$ peak location. 
We have widely discussed in Subsection \ref{preliminaire} the behavior
of ALEPH, BELLE and CLEO residuals under very close conditions\footnote{
The differences between Figures \ref{ResTau_abs}  and \ref{dmdg} is essentially
due to the fact that we now perform a simultanous fit of ALEPH, BELLE and CLEO 
data sets instead of examining each them in isolation. In these conditions
the values found for $\delta g$ and $\delta m^2$ are dominated by BELLE and
CLEO which prefer a smaller value  for $\delta g$ than ALEPH. }. 
Figure (\ref{ResTau_abs_spect}) gives
another view~: Comparing Figure (\ref{ResTau_abs_spect})
to Figure (\ref{ResTau_norm_spect}) shows that the residual distributions remain
flat up to the $\simeq~\phi$ mass, but slightly shifted towards positive values.

The error covariance matrix of the fit parameters still exhibits some large correlations
as $(g, \delta g) \simeq 40 \%$ and
$(g, \delta m^2)\simeq  (\delta g, \delta m^2)\simeq -40 \%$. One
also gets correlations at the +20\% level between $\lambda_{ALEPH}$, $\lambda_{BELLE}$ 
and $\lambda_{CLEO}$, while all others do not exceed a few percent level.

The sign of the correlations for $(g, \delta m^2)$ and $(\delta g, \delta m^2)$
prevents a possible relation of the form 
$m^2 + \delta m^2= a [g+ \delta g]^2 f_\pi^2$ for the $\rho^\pm$ mass squared.
Relying on the global fit of all $\tau$ data samples, one gets~: 
\be
\left\{
\begin{array}{llll}
\displaystyle \frac{\Gamma_{\rho^\pm}-\Gamma_{\rho^0}}{\Gamma_{\rho^0}} \simeq &
\displaystyle \frac{2 \delta g}{ g} =& [-1.10 \pm0.26] ~\%\\[0.5cm]
\displaystyle m^2_{\rho^\pm}-m^2_{\rho^0}  =& \delta m^2=&(1.06 \pm 0.62) ~10^{-3} 
~~{\rm GeV}^2
\end{array}
\right .
\label{rhobrk2}
\ee
significantly smaller than for ALEPH only (see Eq. (\ref{rhobrk})).
In this case, the mass difference between the two $\rho$ mesons is
$m_{\rho^\pm}-m_{\rho^0}\simeq 0.68 \pm 0.39$ MeV, whereas
$\Gamma_{\rho^\pm}-\Gamma_{\rho^0} \simeq -1.63 \pm 0.39$ MeV only.

Summarizing the results gathered in Table \ref{T2},
one can conclude that $e^+ e^-$  data 
and $\tau$ data  do not exhibit inconsistencies. Indeed, the fit of the 
spectra reported  in the third data column is quite satisfactory, and this
takes into account all reported experimental information.
This agreement at the the form factor level,
does not {\it a priori} mean  that similar  $g-2$ values will be obtained
when using/removing  $\tau$ data.

Taking into account the statistical significance of the rescaling factors
as reflected by their uncertainties (see Table \ref{T2}), 
we neglect  none of them in the rest of this analysis. One may stress once again that 
these rescaling factors may well only reflect  an incomplete account of the 
isospin breaking
effects affecting the $\tau$ spectra lineshape which have to be "subtracted"
 in order to compute estimates of the $\pi \pi$ loop contribution  to  $g-2$.

\vspace{0.5cm}

Finally, one may also wonder to reach as frequently such favorable probabilities.
As can be seen in Table \ref{T2}, this reflects
 the low contributions to the $\chi^2$ yielded by the old $\pi \pi$ 
data \cite{Barkov,DM1} ($\chi^2/n \simeq 0.6$)
and by the (recent) $e^+ e^- \ra (\pi^0/\eta) \gamma$ 
data \cite{CMD2Pg1999,CMD2Pg2001,CMD2Pg2005,sndPg2000,sndPg2003,sndPg2007}
($\chi^2/n \simeq 0.75$). Instead, the newly collected  $\pi \pi$  data sets
\cite{CMD2-1995corr,CMD2-1998-1,SND-1998} yield $\chi^2/n \simeq 0.9$ and
the 3--pion data 
\cite{CMD2-1995corr,CMD2-2006,CMD2KKb-1,CMD2-1998,ND3pion-1991,CMD3pion-1989} 
$\chi^2/n \simeq 1.1$, quite comparable to CLEO data $\chi^2/n \simeq 1.3$.
Discarding the contribution of the scale penalty terms, BELLE data yield
$\chi^2/n \simeq 1.4$ and ALEPH $\simeq 0.9$. Therefore, these high probabilities reveal
certainly some overestimation of systematic errors in definite $e^+ e^-$ data sets. When
data sets for $e^+ e^- \ra (\pi^0/\eta) \gamma$  with better estimated
errors will become available, this question will be naturally solved~;
awaiting this, one may conclude that the most precise pieces of 
information provided presently by the full  $e^+ e^- \ra (\pi^0/\eta) \gamma$ 
cross sections are essentially the partial widths for 
$\rho^0/\omg/\phi \ra (\pi^0/\eta) \gamma$.

On the other hand, this explains why one  should carefully control that
some inconsistency in the fit is not hidden by these favorable probabilities.
A good criterium is certainly provided by the $\chi^2/n$ value associated with
each data set and by its behavior while modifying the data set collection
submitted to fit.  

\section{Consequences for the Hadronic Contribution to $g-2$}
\label{gMoins2}
\indent \indent An important outcome of  our model and of our treatment
of the data sets -- {\it i.e.} our fitting procedure -- is related with 
the estimate of hadronic contributions to $g-2$ up to 1 GeV.
Mixing different processes correlated
by the same underlying physics cannot be successful without some clear
understanding of the errors in each data set, which should be 
properly implemented inside the fitting code. 
Our procedure  relies on the whole available information on uncertainties
(magnitude  and type). This allows us to draw conclusions  directly
from a global fit to several data sets  rather than combining 
physics information derived
from separate fits to the individual data samples.

However, this does not guarantee that 
combining  various experiments will result in improved
uncertainties. Indeed, the way the systematic errors, especially scale uncertainties,
combine in the fit procedure cannot be straightforwardly guessed.
One  may, nevertheless, expect the results to be more reliable~; 
indeed, assuming systematics are randomly distributed,
the net result of mixing different experiments and/or data sets should
be to neutralize them to a large extent\footnote{
Of course, one cannot completely  exclude that systematics could "pile up" 
coherently~; however, if the number of different data sets 
is large enough, such a possibility looks rather unlikely.}.

\vspace{0.5cm}

 The lowest order contribution of a given annihilation process 
 $e^+ e^- \ra H$ to the muon anomalous magnetic moment $a_\mu=(g-2)/2$ 
 is given by~:
 \be
 \displaystyle a_\mu (H) = \frac{1}{4 \pi^3} \int_{s_H}^{s_{cut}}  ds ~K(s)~ \sigma(s)
 \label{eqp2}
 \ee
 where $\sigma(s)$ is the Born cross section of the annihilation process
 $e^+ e^- \ra H$, $s_H$  its threshold squared mass and $s_{cut}$
 is an assumed end point of the non--perturbative region.
 $K(s)$ is a  known kernel \cite{Fred09}  given by the integral~:
 \be
 \displaystyle  K(s)=\int_0^1 dx \frac{x^2(1-x)}{x^2+(1-x)s/m_\mu^2}~~~,
 \label{eqp3}
 \ee 
 $m_\mu$ being the muon mass.
 For $s>4 m_\mu^2$, this writes~:
 \be
 \left \{
 \begin{array}{lll}
 \displaystyle K(s)= \frac{x^2}{2}(2-x^2) + \frac{(1+x^2)(1+x^2)}{x^2}
 \left[ \ln{(1+x)} -x + \frac{x^2}{2} \right] + \frac{1+x}{1-x}~x^2\ln{x}\\[0.5cm]
 \displaystyle {\rm with~~~:~~~} x=\frac{1-\beta}{1+\beta} ~~~{\rm and}~~~\beta =\sqrt{1-\frac{4 m_\mu^2}{s}}
 \end{array}
 \right.
 \label{eqp4}
 \ee 
 and, for $0 < s \leq 4 m_\mu^2$, it becomes \cite{Fred09}
 ($r=s/m_\mu^2$)~:
  \be
  \displaystyle K(s)= \frac{1}{2}- r +\frac{1}{2} r (r-2) \ln{r} -(1 -2 r + \frac{r^2}{2})
  \sqrt{\frac{r}{4 -r} }\arctan{\sqrt{\frac{4-r}{r}}}
 \label{eqp5}
 \ee 
This expression has to be used in order to integrate the cross section for  
$e^+ e^- \ra \pi^0 \gamma$ below the two--muon threshold.
 
 Our global fit provides the theoretical Born cross sections with their
 parameter values, errors and their (full) covariance matrix. As illustrated above and in
 \cite{ExtMod1}, the results obtained while fitting altogether several $e^+e^-$  
 cross sections and $\tau$ spectra are satisfactory. Therefore,
 we consider  that using our cross sections  
 within a Monte Carlo, which fully takes into account the parameters, their errors and correlations,
 should provide a  fairly well motivated value
  for each accessible $a_\mu (H)$ in our
 fitting range ({\it i.e. } from each threshold up to $\simeq$1 GeV/c) and for its
 uncertainty which appropriately merges statistical and systematic errors. 
 
 In order to allow for a motivated comparison between our results and experimental 
 estimates for $a_\mu(\pi^+ \pi^-)$,  one should also
  include the effects of Final State Radiation (FSR)
 into our estimates of the $\pi \pi$ contribution to the muon anomalous moment.
 Indeed, even if not always manifest, the shift produced by FSR corrections
 is included in the reported \cite{CMD2-1995corr,CMD2-1998-1,SND-1998} 
 experimental values for $a_\mu(\pi^+ \pi^-)$. The FSR corrections are accounted for
 by multiplying the $\pi \pi$ Born cross section in Eq. (\ref{eqp2}) by \cite{Fred09}~: 
 \be
 \displaystyle 
 1+\eta_{FSR}(s)=
 \left ( 
 1+ \eta(s) \frac{\alpha_{em}}{\pi} - \frac{\pi \alpha_{em}}{2 \beta} 
 \right )
 \frac{\pi \alpha_{em}}{\beta}  
 \left( 
 1-\exp{\{-\frac{\pi \alpha_{em}}{\beta}\}}
  \right )
 \label{FSR}
 \ee
 where the Schwinger function\cite{Schwinger}  $\eta(s)$ can be found 
 corrected for a missprint in \cite{Drees} together with a simplified 
 expression (also derived by Schwinger) valid at the 1.5\% level. The uncertainties
 affecting the FSR effect estimates are not known~; nevertheless, they are expected small in our 
 range of interest\cite{Fred09}~;
 we, therefore, neglect their contribution to the errors we report on $a_\mu(\pi \pi)$.
  Finally, FSR effects on contributions other than  $\pi \pi$ to $a_\mu$ are also known
 to be negligible  \cite{Fred09} and are thus neglected.

 \subsection{Global Fit Results And $e^+e^-$ Data}
 \indent \indent
 It is quite important to compare the outcome of our model and fit procedure with 
 experimental data. This should allow  to check for  possible methodological biases and 
 to substantiate how the merging of statistical and systematic errors operates.
 
 CMD-2 \cite{CMD2-1995corr,CMD2-1998-1} and SND  \cite{SND-1998}, for instance, have
 published  $a_\mu(\pi^+ \pi^-)$ obtained by integrating numerically their measured
  $e^+ e^- \ra \pi^+ \pi^-$ cross section over the interval $\sqrt{s}=0.630 \div 0.958$ GeV,
 using  the trapezoidal method.
  We have run our code using separately each of the data sets given in resp. \cite{CMD2-1995corr},
  \cite{CMD2-1998-1} and \cite{SND-1998}  together with the relevant set of decay partial widths
  of vector mesons needed in order to determine numerically the SU(3)/U(3)/SU(2) breaking
  parametrization. 

\begin{table}[!htb]
\hspace{-2.cm}
\begin{tabular}{|| c  || c  | c | c  | c ||}
\hline
\hline
\hhhu Data Set  & Experimental Result &   Fit Solution & \multicolumn{2}{|c|}{Statistical Information} \\
\hline
\hhhu ~~ & ~~& ~~&  $\chi^2/\rm{dof}$ & Probability\\
\hline
\hline
\hhhu CMD--2 (1995)& $362.1 \pm (2.4)_{stat} \pm (2.2)_{syst}$   &
 $362.57\pm 2.64$ & $42.46/44$ & 53.8\%\\
\hline
\hhhu CMD--2 (1998)& $361.5 \pm (1.7)_{stat} \pm (2.9)_{syst}$    
& $362.36 \pm 2.14$  &$38.05/40$  & 55.9\%\\
\hline
\hhhu  SND (1998)  & $361.0 \pm (1.2)_{stat} \pm (4.7)_{syst}$ 
& $361.09 \pm 2.04$  & $27.10/46$ & 98.8\% \\
\hline 
\hline 
\hhhu  NSK New Timelike $\pi^+\pi^-$ & $360.0 \pm 3.02_{exp}$ $~~^{***}$
& $361.39 \pm 1.72$ &  $124.61/128$ & 56.8\% \\
\hline 
\hhhu  NSK New Timelike~$\pi^+\pi^-$ + KLOE & $358.5 \pm 2.41_{exp}$ $~~^{***}$
& $360.25 \pm 1.48$ &  $255.22/188$ & 0.1\% \\
\hline 
\hline 
\hhhu NSK($\pi^+\pi^-$)  & ~~& $359.50 \pm 1.60$ &  $180.09/210$ &  93.3\%\\
\hline
\hhhu  NSK ($\pi^+\pi^-$) + $(\pi^0/\eta) \gamma$ &  & $359.42\pm 1.52$ & 373.59/468 & 99.96\%\\
\hline
\hline
\hhhu  NSK ($\pi^+\pi^-$) + $(\pi^0/\eta) \gamma$ + ($\pi^+ \pi^- \pi^0$) & 
& $359.31 \pm 1.62 $ &  525.10/597 & 98.4\% \\
\hline
\hline
\end{tabular}
\caption{
\label{T3} Contributions to $10^{10} a_\mu(\pi \pi)$ from the invariant mass region $0.630-0.958$ GeV/c. 
The data flagged by $~~^{***}$ are combined values
proposed in \cite{DavierHoecker} for the New Timelike data altogether, with or without KLOE
data \cite{KLOE08}. The first three lines  in this Table refer
to $a_\mu(\pi \pi)$ experimental values given in  resp. \cite{CMD2-1995corr}, \cite{CMD2-1998-1} 
and \cite{SND-1998}. "NSK ($\pi^+\pi^-$)" indicate that {\it all}  
annihilation data to $\pi^+\pi^-$  are considered.}
\end{table}

  The results are given in the upper part of Table \ref{T3} and look
  in good correspondence with expectations. The highly favorable probability for the
  SND spectrum \cite{SND-1998} might be attributable to its larger systematics compared to CMD--2.
  The middle line in this Table shows the result obtained while fitting simultaneously
  the three Novosibirsk spectra \cite{CMD2-1995corr,CMD2-1998-1,SND-1998}  together. The global fit
  of these data sets, where systematics  are certainly under good control, 
  provides a strong reduction of the  global uncertainty -- merging all reported systematic 
  and statistical errors. The improvement of the  uncertainty derived from the global fit solution
  compares favorably  with the average value proposed by \cite{DavierHoecker} using  a spline
  method.   Therefore,  nothing obviously abnormal is recognized in this comparison.
  The following line shows that KLOE data \cite{KLOE08} allows to reduce
  a little bit more the uncertainty~; both the central value and the uncertainty
  compare well with the average proposed by \cite{DavierHoecker}. However, one should note
  the very low probability of the associated global fit.

 The next three lines in Table \ref{T3} are also quite interesting. The first of these
 reports the result obtained when fitting using all available $\pi^+ \pi^-$ data 
 sets -- except for KLOE \cite{KLOE_ISR1}~; this turns out to include into the fitted data the
 so--called "Old Timelike Data" \cite{Barkov,DM1} (82 measurements) besides the "New 
 Timelike Data" (127 measurements).
 The central value for $a_\mu(\pi \pi)$  is slightly shifted downwards by
 about $1 \sigma$ with a negligible gain in precision. One also observes that the fit 
 probability makes a large jump ($\simeq 50 \% \Rightarrow \simeq 90 \%$), reflecting the larger 
 uncertainties in the data from \cite{Barkov,DM1} compared to those from 
 \cite{CMD2-1995corr,CMD2-1998-1,SND-1998}.  This once more shows that the
 systematics in the data sets from  \cite{Barkov,DM1} have been conservatively estimed.
 
 The last 2 lines in Table \ref{T3} exhibit the same trend~; 
 indeed, including in the fit procedure the 86 measurements
 of $e^+e^- \ra \pi^0 \gamma$, the  182 measurements of $e^+e^- \ra \eta \gamma$
 and the 126 points of  $e^+e^- \ra \pi^0 \pi^+ \pi^-$ from 
 \cite{CMD2-1995corr,CMD2-2006,CMD2KKb-1,CMD2-1998,ND3pion-1991,CMD3pion-1989}
 does not improve, 
 strictly speaking,  the uncertainty~; however, it is satisfactory to check
 that the central value only fluctuates inside quite acceptable limits. 
 This teaches us some remarkable facts~: 
 
  \begin{itemize}
 \item
 Improving systematics in all processes having the same 
 underlying physics as $\pi^+ \pi^-$ might be useful 
 to allow a better determination of their own contributions to $a_\mu$, but also of
  $a_\mu(\pi \pi)$ itself. Indeed, most physics parameters in the quoted processes
  are the same as in $e^+ e^- \ra \pi^+ \pi^-$. Conversely, higher quality
  $\pi^+ \pi^-$ data should improve estimating the contributions  of the 
  other related annihilation processes to $a_\mu$.

\item If errors are reasonably well understood and appropriately dealt with, adding
poor quality data sets into the fitted sample does  not degrade the result~;
this allows, however, to confirm the stability of the central values,
which is certainly a valuable piece of information.

\end{itemize}

However, the most important remark is certainly that using the radiative partial width 
decays of light mesons -- and/or the $e^+ e^- \ra (\pi^0/\eta) \gamma$ cross sections -- together
with $e^+ e^- \ra \pi^+ \pi^-$  allows a quite significant improvement of the
accuracy on $a_\mu(\pi \pi)$ compared to the direct numerical integration of the
experimental spectra. This is, indeed, the main advantage of having a constraining global model 
allowing for an overconstrained  global fit.

\subsection{Effects Of Including $\tau$ Spectra And KLOE Data}
\indent \indent One may question the effects produced by introducing 
the $\tau$ decay data inside our collection of fitted data sets. These 
are expected to improve the model parameter values (and errors), if their systematics 
are indeed reasonably well controlled. 

Our strategy will be to consider all $e^+ e^-$ data sets 
used just above and look at the effect of including the A \cite{Aleph}, B \cite{Belle} and
C \cite{Cleo} data samples in isolation or combined.

As above, when using the fit results obtained by
constraining the rescaling factors -- {\it i.e.}  keeping the $(\lambda/\eta)^2$ terms
in the $\chi^2$ expressions for $\tau$ data samples -- the $\tau$ samples will be denoted
A, B and C~; when concentrating over $\tau$ lineshapes, these will be
denoted A$^{sh}$, B$^{sh}$ and C$^{sh}$. The case when fitting the $\tau$ spectra
by allowing non--vanishing $\delta g$ and $\delta m^2$ and imposing $\lambda \equiv 0$
will still be referred to as
A$_{dm,dg}$ B$_{dm,dg}$  or C$_{dm,dg}$ in our Tables and/or Figures.

\subsubsection{Comparison With Standard $\tau$ Based Estimates for  $a_\mu$}
\indent \indent When letting free $\delta g$ and $\delta m^2$  while imposing
 $\lambda \equiv 0$, our approach is  comparable  to those underlying
 the so--called $\tau$ based estimates \cite{DavierHoecker} of the hadronic contribution 
 to the muon $g-2$.
 Indeed, the  whole set of $e^+ e^-$ data fixes in a data driven mode all identified isospin
 breaking corrections~:  $\rho-\omg$ and $\rho-\phi$ meson mixing, $\rho$ meson mass
 and width differences, I=0 part of the $\rho$ revealed by its coupling to 
 photon\footnote{This contribution is currently not considered as such \cite{DavierHoecker}.
 It should be partly absorbed in $\rho$ meson width corrections.
 } (see Eq. (\ref{referee1})) and FSR corrections. The pion mass difference is plugged in 
 directly.  It is thus interesting to compare our results in this case with the existing estimates
 and thus check the effects of a global fit. 
The hadronic contributions to the muon $g-2$  derived from $\tau$ data
have been updated recently in \cite{DavierHoecker}~;
the numerical contributions to $a_\mu$ from the the reference region $\sqrt{s} \in [0.63,0.958]$
are  not published but have been kindly communicated to us \cite{ZhangPriv}. 

\begin{table}[!htb]
\hspace{-2.cm}
\begin{tabular}{|| c  || c  | c | c | c | c ||}
\hline
\hline
\hhhu Data Set  &  $a_\mu(\pi\pi) $ &  $a_\mu(\pi\pi) $    & \multicolumn{3}{|c|}{Statistical Information} \\
\hline
\hhhu ~~ & Experimental Result & Fit Solution &  $\chi^2_\tau/\rm{dof}$&  $\chi^2_{ee}/\rm{dof}$ & Probability\\
\hline
\hline
\hhhu A$_{dm,dg}$ \cite{Aleph}& $364.02 \pm (2.19)_{exp} \pm (1.97)_{Br} \pm (1.51)_{IB}$   &
 $362.83\pm 1.46$ & $45.52/37$ & $122.12/127$ & 95.3\%\\
~~& $(364.02 \pm 3.31_{tot})$ & ~~  & ~~& ~~& ~~\\
\hline
\hhhu  B$_{dm,dg}$ \cite{Belle} &    $366.44 \pm (1.02)_{exp} \pm (5.70)_{Br} \pm (1.51)_{IB}$  
& $364.85 \pm 1.32$  &$41.95/19 $  &$128.58/127 $ & 76.5\%\\
~~& $(366.44 \pm 5.98_{tot})$ & ~~  & ~~& ~~& ~~\\
\hline
\hhhu  C$_{dm,dg}$  \cite{Cleo} &  $366.62 \pm (4.17)_{exp} \pm (6.37)_{Br} \pm (1.51)_{IB}$  
& $364.23 \pm 1.82$  & $63.01/29$ & $125.63/127$& 70.0\% \\
~~& $(366.62 \pm 8.05_{tot})$ & ~~  & ~~& ~~& ~~\\
\hline 
\hline 
\hhhu  OPAL  \cite{Opal} &  $354.40 \pm (4.67)_{exp} \pm (4.78)_{Br} \pm (1.51)_{IB}$  
& --  &  -- &  -- & --  \\
~~& $(354.40 \pm 6.85_{tot})$ & ~~  & ~~& ~~& ~~\\
\hline 
\hline 
\hhhu ALL $\tau$ Sets  &   $367.46 \pm (1.31)_{exp} \pm (1.59)_{Br} \pm (1.51)_{IB}$ 
&   $367.12 \pm 1.30$  &$102.06/85 $&$143.32/127$ &  58.2\%\\
~~& $(367.46 \pm 2.55_{tot})$ & ~~  & ~~& ~~& ~~\\
\hline
\hline
\end{tabular}
\caption{
\label{T3b} Contributions to $10^{10} a_\mu(\pi \pi)$ from the invariant mass region $0.630-0.958$ GeV/c. 
The experimental values \cite{ZhangPriv} are derived using the method from \cite{DavierHoecker}. 
The experimental average includes OPAL data, our fit result does not.
The meaning of  A$_{dm,dg}$, B$_{dm,dg}$, C$_{dm,dg}$ is explained in the text.}
\end{table}

Table \ref{T3b} displays the experimental data derived from each existing $\tau$ data set
and their combination \cite{DavierHoecker,ZhangPriv} in the first data column. One may note
that the proposed experimental average is larger than each of the individual estimates~; 
actually, in order
to perform the average, each individual estimate has been rescaled to the world
average value for ${\cal B}_{\pi \pi}$ \cite{ZhangPriv}. The total (experimental)
errors displayed  are computed by summing up the various components in quadrature. This is
provided in order to allow for an easy comparison with our fit result.
The data columns $\chi^2_\tau/\rm{dof}$ and $\chi^2_{ee}/\rm{dof}$ display 
the contribution to the total $\chi^2$ provided by resp. the $\tau$ and the so--called New Timelike
$e^+ e^-$ annihilation data \cite{CMD2-1995corr,CMD2-1998-1,SND-1998}, which serve as 
quality tags.

One should note the nice correspondence between our central values and the corresponding 
experimental estimates. The improvement of the total errors provided 
by the global fit method is also worth mentioning. The reduction of the uncertainties
provided
when using $\tau$ data looks even more important that when using $e^+e^-$ data alone.
Of course, the errors provided there, as anywhere in this paper, are the {\sc minuit} errors 
returned by the fits.

The result for (A$_{dm,dg}$ B$_{dm,dg}$ C$_{dm,dg}$) -- last line in Table \ref{T3b} -- 
is also quite remarkable. It clearly 
illustrates that our global fit does not lead to a standard averaging, but takes into account
the relative mismatch of the A, B and C lineshape distortions noted in Subsection \ref{preliminaire}. 
In this procedure, the fit average 
is significantly pushed upwards, in accord with  the experimental estimate\footnote{
This should be slightly larger, if removing the OPAL  data set \cite{Opal}.}.  

The data columns providing the  $\chi^2$ information are also quite important.
As already noted in Subsection \ref{preliminaire},  $\chi^2_{ALEPH}$ is reasonably good
simultaneously with the New Timelike $e^+ e^-$ data. Comparing  $\chi^2_{ALEPH}$ here
(45.52) with its homologue in Table \ref{T0} -- fourth line therein -- (29.17), reveals that
the ALEPH data sample  meets some difficulty in accomodating the data for
$e^+ e^- \ra \pi^+ \pi^- \pi^0 $. However, the description remains quite reasonable and
one may conclude that there is no real mismatch, within our approach,  between ALEPH data
and the VMD expectations. The  $a_\mu$ value just derived from ALEPH data compares
well with those derived from $e^+ e^-$ data only (see Table \ref{T3}).

The values for $a_\mu$ derived for BELLE and CLEO data in isolation, even if slightly
larger than expected from VMD inference, are not in  strong disagreement with these.
However, as can be seen from either of Table \ref{T3b}
 and Table \ref{T0} (see the $\lambda \equiv 0$ entries therein), the fit quality, as
 reflected by $\chi^2_{BELLE}$ and $\chi^2_{CLEO}$ looks significantly
 poorer (both yield $\chi^2/npoints \simeq 2$) than  ALEPH 
 ($\chi^2/npoints \simeq 1.1 \div 1.2$). Whether this behavior 
 reveals specific systematics is an open issue.

When performing the global fit with all $\tau$  data samples (last line in Table \ref{T3b}), 
the weight of $\tau$ data happens to become dominant. 
In this case, the fit of the $\tau$ data is nearly unchanged
($\chi^2_{ALEPH}=26.64$, $\chi^2_{BELLE}=45.89$, $\chi^2_{CLEO}=29.52$)~; however,
even if apparently reasonable, the new timelike $e^+ e^-$ data yield a $\chi^2_{ee}$
increased by 20 units, which is significantly far enough from optimum that we consider
the corresponding $a_\mu$ value with great care. 

The analysis in Subsection \ref{preliminaire} has shown that  rescaling factors and
lineshape distortions are sharply related. However, in order to derive 
$\tau$ based estimates for $a_\mu$, one should remove all distortions produced by
isospin breaking effects specific of the $\rho^\pm$ meson relative to $\rho^0$.
Therefore, we do not consider reliable the estimates provided in  Table \ref{T3b},
except possibly those in the A$_{dm,dg}$ entry, as the distortions in the B and C
samples are to be better understood.  

\subsubsection{Additional $\tau$ Based Estimates of $a_\mu$}
\indent \indent Our analysis in Subsection  \ref{preliminaire} provided a serious
hint that rescaling the $\tau$ spectra can be an effective way to account for
shape distortions produced by IB effects which differentiate  the $\rho^\pm$ 
and $\rho^0$ mesons. We just provided results assuming no scale correction
to the  $\tau$  data samples ($\lambda_{ALEPH}=\lambda_{BELLE}=\lambda_{CLEO}=0$). 
In order to provide a complete picture, it is worth considering two more sets of fitting conditions.
In this Subsection, we will present the results derived by~:
{\cal i/} fitting the $\tau$ spectrum lineshapes by the method emphasized in Subsection
\ref{taulineshape},  {\cal ii/} fitting the $\tau$ spectra
by allowing non--vanishing rescalings, however,  constrained by the relevant
$(\lambda/\eta_{Exp})^2$ term for each of the $\tau$ data set.

\begin{table}[!htb]
\hspace{+1.5cm}
\begin{tabular}{|| c  ||  c | c  | c ||}
\hline
\hline
\hhhu Data Set  &   Fit Solution & \multicolumn{2}{|c|}{Statistical Information} \\
\hline
\hhhu ~~ & $10^{10} a_\mu(\pi \pi)$ &  $\chi^2/\rm{dof}$ & Probability\\
\hline 
\hhhu NSK  ($e^+ e^-$)  & $359.31 \pm 1.62$ &  $525.10/597$ &  98.4\%\\
\hline 
\hline 
\hhhu NSK +A$^{sh}$ B$^{sh}$ C$^{sh}$   & $359.62 \pm 1.51$ &  $605.90/679$ &  97.9\%\\
\hline 
\hline 
\hhhu NSK   + A  ($\chi^2_\tau=$ 28.39)   &  $362.24 \pm 1.52$ & 568.60/632   &  96.6\% \\
\hline
\hhhu  NSK  + B  ($\chi^2_\tau=$ 32.59)    &  $360.68 \pm 1.47$ & 558.88/614 &  94.6\% \\
\hline
\hhhu NSK   + C  ($\chi^2_\tau=$ 39.13)   &  $360.52 \pm 1.55 $ & 565.24/624  &  95.5\% \\
\hline
\hhhu  NSK  + A B C   &  $364.48 \pm 1.34$ &  648.60/680   &  80.1 \% \\
\hline
\hline
\hline 
\hhhu NSK   + KLOE    & $359.22 \pm 1.32$ &652.92/657& 51.9\%\\
\hline
\hhhu NSK  + KLOE + A$^{sh}$ B$^{sh}$ C$^{sh}$  & $358.52 \pm 1.32$ & 741.50/739 & 46.7\% \\
\hline
\hhhu NSK  + KLOE + A B C   & $364.04 \pm 1.25$ & 792.28/740 & 8.9\% \\
\hline
\hhhu NSK  + KLOE + A  ($\chi^2_\tau=$ 49.59)   & $361.55 \pm 1.31$ & 708.90/692 & 32.0\% \\
\hline
\hhhu NSK  + KLOE + B ($\chi^2_\tau=$ 34.06) & $360.19 \pm 1.16$ & 688.02/674 & 34.6\% \\
\hline
\hhhu NSK  + KLOE + C ($\chi^2_\tau=$ 39.89)& $360.11 \pm 1.32$ & 693.40/684 & 39.3\% \\
\hline
\end{tabular}
\caption{
\label{T4} Contributions to $10^{10} a_\mu(\pi \pi)$ from  $\sqrt{s} \in[0.630,0.958]$ GeV/c. 
$\tau$ data set configurations are considered together with the $e^+e^-$ data and the appropriate 
subset of decay partial widths. The meaning of A, B, C, and 
A$^{sh}$, B$^{sh}$, C$^{sh}$ is given in the text. The contribution of single $\tau$ subsets
to the total $\chi^2$ is indicated when relevant by ($\chi^2_\tau=\cdots$). For lineshape 
fits without KLOE data,
we have $\chi^2_\tau=$ 18.22 for A, $\chi^2_\tau=$ 28.35 for B and $\chi^2_\tau=$ 31.88 for C. } 
\end{table}

The upmost part of Table \ref{T4} collects our  results for $a_\mu$ obtained using the $\tau$ data samples, 
except for the first line which is nothing but the final result in Table \ref{T3}. The second 
line shows that including the $\tau$ lineshapes into the fit (A$^{sh}$, B$^{sh}$, C$^{sh}$) gives results 
perfectly consistent with using only $e^+ e^-$ data and a slightly improved uncertainty.
As the fit quality of each   A$^{sh}$, B$^{sh}$ and C$^{sh}$ on the one hand, and each of the 
$e^+ e ^-$ data are simultaneously optimum, one may consider this estimate safe. 

Compared with pure $e^+e^-$ estimates, the next 3 lines show the effects of including each 
of A, B and C in  isolation\footnote{And keeping the  $(\lambda/\eta)^2$ term included in the 
$\chi^2$ expression, as stated above.}.  The $a_\mu$ value found for ALEPH agrees with that
in the A$_{dm,dg}$ entry in Table \ref{T3b}. Interestingly, the values found using either 
B or C are significantly lower than their homologues in Table \ref{T3b},
but in nice correspondence with VMD expectations (see Table \ref{T3}). This may  
indicate that some rescaling is needed for these, consistent with their uncertainties.

A simultanoueous use of A, B and C (line flagged with NSK+ABC) provokes
an upward shift by    $4~10^{-10}$, as already observed in  Table \ref{T3b}.
The behavior of the combination under the fit
illustrates, once again,  
that the shape distortions exhibited by each of A, B and C are
 different and that the compensation performed by the fit
degrades the description of the $e^+ e^-$ data. This produces  the
increased value for $a_\mu$, as also observed in the previous Subsection.

\subsubsection{Including KLOE Data}
\indent \indent 
 Up to now, we have not introduced the KLOE \cite{KLOE_ISR1} data into the data
 sets submitted to a fit. Indeed, even if considered acceptable, its  best $\chi^2/n_{points}$
 looks large \cite{ExtMod1}. It is nevertheless instructive
 to point out explicitly  its effects. Table 2 in \cite{ExtMod1} indicates
 that three among the rescaling coefficients can be cancelled out~; one only has to let
 vary those corresponding to the global scale ($\varepsilon_0$) and to the acceptance
 correction ($\varepsilon_2$) \cite{KLOE_ISR1}. 
 
 This provides the results  given in the lower half
 of Table \ref{T4}. Even if the probability decreases due to the intrinsically high
 (minimum) $\chi^2$ of the KLOE data sample, one observes a good consistency of 
 the $a_\mu(\pi \pi)$  value derived from fitting the (rescaled) KLOE data together with all 
 $e^+ e^-$ Novosibirsk data.  One also starts getting much improved uncertainties.
 One may note the good probability when using all NSK data
 together with KLOE and  A$^{sh}$ B$^{sh}$ C$^{sh}$ and the improved uncertainties on  $a_\mu(\pi \pi)$.
 When using A, B and/or C , one finally observes the same trend as in the upper part
 of Table \ref{T4}, which lead us to be as cautious in this case.
  
  Nevertheless, one may conclude that KLOE data do not degrade the expectations
  from NSK data, whatever is done with $\tau$ data.
  
  In order to assess the effects of a global fit, it is also interesting to compare
 the result for NSK+KLOE in Table \ref{T4} with the corresponding average published
 in \cite{DavierHoecker}~:
 $${\rm~~Global Fit~:}~a_\mu(\pi \pi)= 359.22 \pm 1.32 ~~\Longleftrightarrow
 {\rm Average\cite{DavierHoecker}~:}~a_\mu(\pi \pi)= 358.51 \pm 2.41_{exp} $$
 in units of $10^{-10}$. As for NSK data alone, the central value from our fit is slightly
 larger than the estimate  from \cite{DavierHoecker}
 and the uncertainty is significatively improved.

 \subsubsection{A Partial Summary}
 \indent \indent
In summary, the picture exhibited when using the $\tau$ data samples looks somewhat
intricated. We have shown above, especially in Subsection \ref{preliminaire}, that lineshape 
distortions and absolute scale of spectra are intimately related.

However, the present analysis tends to indicate  that an important part  of
 the reported discrepancy between $e^+ e^-$ and $\tau$ based estimates of $g-2$
 is related with a more or less difficult way to account for isospin breaking effects
 differentiating the $\rho^\pm$ and $\rho^0$ lineshapes.  This conclusion
 is enforced by comparing the behavior of the ALEPH data set with those of BELLE and CLEO~:
the distortions of the former set can be accounted for within
 our model, providing a reasonable accord with  VMD expectations.
 
  As for BELLE and CLEO, genuine mass and width differences compared to $\rho^0$,
  which successfully work with ALEPH data, do not avoid a residual distortion effect
  which has important consequences while estimating $a_\mu$.
 
 Interestingly, the normalized $\tau$ spectra do not reveal the same problem. This
 is certainly due to the fact that a free rescaling  
 allows to disconnect lineshape  distortions as parametrized
 by mass and width differences ($\delta g$ and $\delta m^2$ in our model) from the
 absolute scale. This indicates that the distortions exhibited by BELLE and CLEO
 are not understood. Of course, one cannot exclude that the agreement with ALEPH
 is purely accidental and that a more refined IB mechanism might have to be considered. 
 
\begin{figure}[!ht]
\vspace{0.5cm}
\begin{center}
\includegraphics*[angle=0,width=1.\columnwidth]{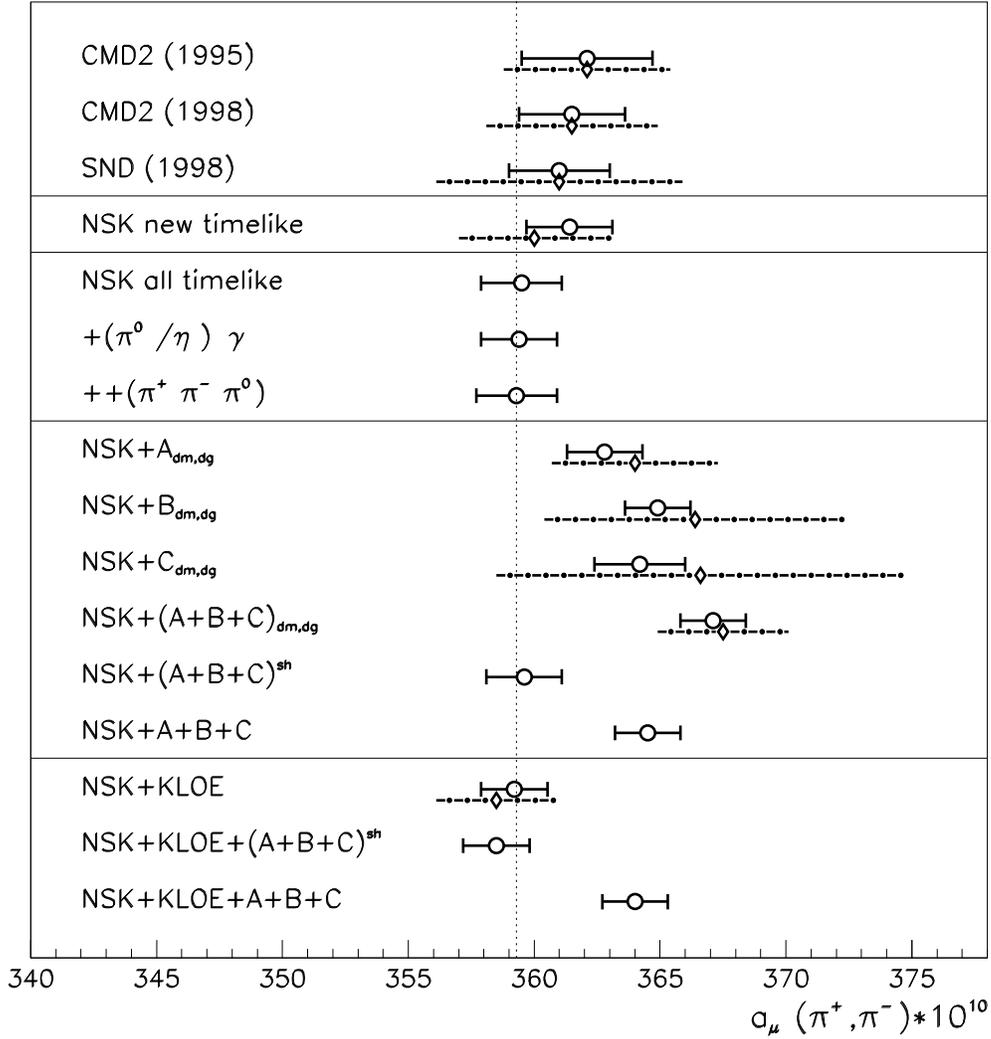}\\
\caption{\label{Compilation}  Values of $a_\mu(\pi \pi)~\times 10^{10}$
integrated between 0.630 and 0.958 GeV.
Data configurations have been defined in Table \ref{T3} for the first seven lines,
and in Tables \ref{T3b} and \ref{T4} for the following lines. The various notations
used for the $\tau$ data samples have been defined in the text. The points with 
dashed--dotted uncertainties are experimental values provided by the experiments
 \cite{CMD2-1995corr,CMD2-1998-1,SND-1998} or in \cite{DavierHoecker}. 
}
\end{center}
\end{figure}

  Figure \ref{Compilation} displays graphically the main conclusions discussed in this Section.
   The vertical line centered at the value derived by fitting all Novosibirsk data is drawn
   to guide the eye. Comparing our estimates with the experimental data reported there is also 
   interesting by itself. This indicates that individual estimates of $a_\mu$ provided
   by each of the  $\tau$ data sets are close to the $e^+e^-$ based estimate.
   The combined  fit of these, instead, exhibits a clear discrepancy, as shown by the data points
   flagged   by NSK + A+B+C , NSK + A$_{dm,dg}$+B$_{dm,dg}$+C$_{dm,dg}$ or with KLOE. 
   Interestingly, the comparison of the various data points exhibits the same trend.

 \subsubsection{Comments On Our Model Uncertainties}
 \label{moderr}
 \indent \indent 
 Uncertainties intrinsic to our model may have to be estimated. For this purpose, 
 having  alternative models able to cover the same scope than our extended model  
 would be desirable. Unfortunately, such models do not seem to exist which could  cope with
 detailed descriptions of several decay channels over our whole fitting range.
 Some interesting attempts have been initiated relying on Chiral Symmetry and the properties of the
 Roy equations \cite{Colangelo1,Colangelo2}~; however, nothing final is presently available.
 
 In our case, one cannot make estimates by changing, for instance. the $\rho$
 parametrization (Gounaris--Sakurai, Breit--Wigner, Kuhn--Santamaria \ldots) as sometimes done
 elsewhere. Indeed, the form factor lineshape is intrinsic to the model, including the
 IB vector meson mixing. However, the difference between the central values of our estimates  
 and the corresponding experimental estimates
 \cite{CMD2-1995corr,CMD2-1998-1,SND-1998} may give some hint on the bound
 for our model uncertainty.

\subsection{Comparison with  $a_\mu$ Estimates Collected by the ISR Method}
\indent \indent
As a summary, our preferred final result is derived using all $e^+e^-$ 
data collected at Novosibirsk~:
\be
a_\mu(\pi \pi;~0.630\div0.958 {\rm~~GeV}) = [
359.31 \pm1.62_{exp} ]~ 10^{-10} 
\label{finalamu} 
\ee
and corresponds to an (underlying) fit probability of 98.4\%. 

This result can be compared with the recent estimate published by KLOE \cite{KLOE08}~:
  $$a_\mu(\pi \pi; 0.630 < m_{\pi\pi}<0.958) = (356.7 \pm 0.4 \pm 3.1)~~ 10^{-10}$$
based on a newly collected data set \cite{KLOE_ISR1,KLOEnote}. This estimate
may look a little bit low, but is consistent with ours at $\simeq 0.7 ~\sigma$
level. 

On the other hand, the BaBar Collaboration has recently finalized a new data 
set \cite{BaBar}, also collected using the ISR method. The contribution it
provides  to $a_\mu(\pi \pi)$ in the canonical $s$ interval can be
found in \cite{DavierHoecker2}~:
$$a_\mu(\pi \pi;~0.630\div0.958 {\rm~~GeV}) = (365.2 \pm 1.9 \pm 1.9)~ 10^{-10} 
=~(365.2 \pm 2.7_{tot})~ 10^{-10}$$
which supersedes a preliminary result presented at the Novosibirsk Conference 
TAU08\cite{Davier2009} ($a_\mu(\pi \pi) = (369 \pm 0.8 \pm 2.9)~ 10^{-10}$).
Even if  larger than Eq. (\ref{finalamu}), this estimate is consistent with ours
at the $1.9 ~\sigma$ level. 

Comparing our result with these recent estimates clearly illustrates the
advantage of introducing information beyond the $e^+ e^- \ra  \pi^+ \pi^- $
cross section and a framework which encompasses annihilation processes and
partial width decays of light mesons. Indeed, the IB schemes needed can
be calibrated consistently on the data set.
The fact that the HLS framework does not
presently go  much beyond the $\phi$ mass region is certainly a handicap,
but does not prevent interesting improvements. 

\subsection{Contributions to $a_\mu$ Up To 1 GeV}
\indent \indent
The discussion  above leads us to  consider the best motivated results for $a_\mu$ as
 derived from   fit to the Novosibirsk $e^+ e^-$ data samples
  taking into account the photon hadronic VP (see Eq. (36) in\cite{ExtMod1}).
 A challenging choice
 might be to include the $\tau$ {\it normalized} spectra, which have been shown
 to yield a satisfactory description simultaneously with all   $e^+ e^-$ data.
 We have seen that BELLE and CLEO seem to meet problems with their shape distortions
 which are not clearly identified. However, as A$_{\delta m,\delta g}$ is reasonably
 well understood, one may consider its results for $a_\mu$.
 
 Using these data samples, one can estimate  several
 contributions\footnote{Actually, one could have produced, as well,
 these contributions up to about 1.05 GeV/c in order to include the $\phi$ region.
 As our fitted region includes the $\phi$ peak, our results would be as reliable.
 However, the largest ($K \overline{K}$) contribution involving the $\phi$ region is 
 presently left aside because of the issue raised by \cite{BGPter} and still unsolved.  }
  to $ a_\mu$ up to 1 GeV. They are given in Table \ref{T6}.
  We thus provide the results obtained while fitting without any $\tau$ sample
 in the first data column. The last data column, instead, shows the 
 influence of KLOE data \cite{KLOE_ISR1,KLOEnote}. 
 
  \begin{table}[!htb]
\hspace{-1.5cm}
\begin{tabular}{|| c  || c | c |c |c ||}
\hline
\hline
Process  \hhhu  &  NSK (no $\tau$)  & NSK  + A$^{sh}$ B$^{sh}$ C$^{sh}$ &  NSK  + A$_{\delta m,\delta g}$ & 
NSK  + KLOE + A$^{sh}$ B$^{sh}$ C$^{sh}$\\
\hline
\hhhu   $\pi^+ \pi^-$ & $492.02 \pm 2.24 $  & $ 491.66 \pm 1.98$  
& $496.20\pm 1.85$ & $492.24\pm 1.79$\\
\hline
\hhhu   $\pi^0 \gamma$  & $4.53\pm 0.04 $  & $4.53 \pm 0.04$ 
& $4.55 \pm 0.04$& $4.51 \pm 0.05$\\
\hline
\hhhu  $\eta \gamma$  & $0.17 \pm 0.01 $  & $0.17 \pm 0.01$ 
& $0.17 \pm 0.01$ & $0.17 \pm 0.01$\\
\hline
\hhhu   $\eta^\prime \gamma$  & $0.01 \pm 0.00$  &$0.01 \pm 0.00$ 
&  $0.01 \pm 0.00$ &  $0.01 \pm 0.00$\\
\hline
\hhhu  $\pi^+ \pi^-\pi^0$ & $36.99 \pm 0.55 $  &  $36.97 \pm 0.56 $ 
& $36.83 \pm 0.56$ & $37.07 \pm 0.55$  \\
\hline
\hline
\hhhu  Total & $533.72 \pm2.32$ & $533.34 \pm 2.07 $
& $537.76 \pm 2.08$ & $534.01 \pm 1.90$\\
\hline
Fit Probability & 98.4\%   &98.0\% &  95.3\% & 50.8\%\\
\hline
\hline
\end{tabular}
\caption{
\label{T6} Contributions to $10^{10} a_\mu$ from thresholds up to 1 GeV/c
for various processes fitting the data sets indicated on the first line~;
by NSK we mean the set of all $e^+ e^-$ data except for KLOE.
 The  errors provided merge  the reported statistical and systematic uncertainties.
}
\end{table}
 
 One can note that the estimate based only on $e^+e^-$ annihilation data
 is consistently improved by including within the fitted data sets the $\tau$
 lineshapes from ALEPH, BELLE and CLEO. Instead, using  A$_{\delta m,\delta g}$
 produces a shift by 4 units, while reducing the uncertainty in the same way
 than the $\tau$ lineshapes.
 Whether this estimate should be prefered, is an open question, awaiting
 a better understanding  of the $\tau$ spectrum distortions.  On the other hand,
 including KLOE data  does not provokes differences with using only
 the Novosibirk data, but instead provides improved uncertainties.
  
 The always (negligible) contribution of the annihilation process $\eta^\prime \gamma$  
 is a prediction which is entirely determined by the  $\eta \gamma$
 decay channel and our model which implies a tight correlation between 
 the  $\eta \gamma$ and $\eta^\prime \gamma$ final states \cite{ExtMod1}.
 So close to its threshold, this contribution could have been expected small~;
  it should become obviously larger while including the $\phi$ region.
 
  As a final remark, one  may conclude that the global fit method
  indeed performs as expected.   Moreover, as far as we know, 
  our method is the first one which can associate a probability to the various
  reported contributions to the muon anomalous magnetic moment. Of course, 
  the quoted probabilities refer to the fits underlying the estimates
  and not the estimates themselves, as these are only derived from 
  computations using the fit results (parameter values and full error covariance matrix).
  One may infer that our prefered numerical results for $a_\mu$ -- any of the first
  two data columns in Table \ref{T6} -- should increase the 
  discrepancy between the prediction and the BNL measurement of the muon $g-2$.
  Improved estimates can be expected  from using also the new  data samples 
  based on the ISR method collected by KLOE and BaBar.

\section{Summary And Concluding Remarks}
 \label{FinalConclusion}
 \indent \indent
 In the first part of this study \cite{ExtMod1}, we defined  the Extended
 HLS Model and reminded the mechanisms implementing U(3)/SU(3)/SU(2)
 symmetry breaking, noticeably the vector meson mixing provided by breaking Isospin symmetry. 
 This was shown to
 provide a satisfactory simultaneous description of all $e^+ e^-$  data sets considered.
 The annihilation channels successfully covered ($\pi^+ \pi^-$, $\pi^0 \gamma$, 
 $\eta \gamma$  and $\pi^0 \pi^+ \pi^-$)  represent a large amount of data.
 It is also the largest set of annihilation channels  simultaneously
 analyzed so far.  The present work confirms that the dipion spectrum
 in $\tau$ decay is also in  the reach of this model. 
 
 The $\tau$ data sets collected by the ALEPH, BELLE and CLEO Collaborations 
 have been  carefully examined in isolation and combined within the context of a global fit
 performed together with a large set of  $e^+ e^-$ annihilation data. For this purpose,
 we found appropriate to introduce additional Isospin breaking (IB) effects providing
 distortions of the $\rho^\pm$ lineshape compared to $\rho^0$. This was partly
 considered in our former study \cite{taupaper}, but totally ignored in preliminary versions 
 of the present work like \cite{Pekin}. This additional mechanism (IB shape distortions) 
 consists of a coupling difference $\delta g$ of the   $\rho^\pm$ and $\rho^0$ to a pion pair
 and a mass difference $\delta m^2$ between these two mesons.
 
 Each of the ALEPH (A), BELLE (B) and CLEO (C) experiments reports on a global 
 scale uncertainty for their dipion spectra. This scale uncertainty is much smaller for 
 A (0.51\%) than for B (1.53\%) or C (1.75\%). The absolute scale of each 
  $\tau$ spectrum can then be modified by a factor $1+ \lambda_{A/B/C}$, including a fit parameter 
  constrained, for each
 of A, B and C, by its scale uncertainty. However, it has been proved that, 
 in minimization procedures, 
 $\delta g$, $\delta m^2$ and these scales are tightly correlated. Therefore, one cannot
 interpret  $\lambda_{A/B/C} \ne 0$ as purely reflecting
biases on the measured branching ratios ${\cal B}(\tau \ra \pi^\pm \pi^0 \nu)\equiv {\cal B}_{\pi \pi}$.
To be explicit, the connection between pure lineshape distortion parameters and
absolute scale, which is the subject of Subsection \ref{preliminaire}, 
prevents to tag reliably an undoubtful numerical value for some rescaling.
  
It has been shown -- see Table \ref{T0} -- that the ALEPH data are in accord with VMD expectations, provided 
some shape distortions  ($\delta g$, $\delta m^2$) are implemented. In this case,
$\lambda_A=0$ is even found quite acceptable. This means that, relying on ALEPH
data only, the presently accepted value for ${\cal B}_{\pi \pi}$ is not contradicted
by our analysis. This result is tightly connected with the fact that the lineshape
distortions of the ALEPH spectrum is well accounted for by our parametrization
($\delta g$, $\delta m^2$).

The picture is not that clear with BELLE  and CLEO, indicating that some residual 
problem survives, mostly visible in these data sets~; this could
well be due to having a too simple--minded shape distortion mechanism~; however,
this could also reflect specific systematics in these experiments or, as well,
a real physical problem revealed by their larger statistics. A rescaling 
of their ${\cal B}_{\pi \pi}$ downwards is found, however,  in reasonable accord with 
their reported uncertainties.

Moreover, the {\it lineshapes} of the normalized spectra provided by A, B and C happen to 
exhibit no problem at all when fitted together with the largest possible set of $e^+ e^-$  
annihilation data. This is certainly due to having dropped out the correlation
between absolute scale and lineshape distortions. 
In this case, one even finds no need to introduce significant IB shape distortions.
The $\tau$ lineshapes and the $e^+ e^-$  are optimally fitted as clear from
the individual $\chi^2$ given in Table \ref{T2}.

Collecting all pieces of information in this study, one cannot consider
the existence of a fundamental issue, at the  pion form factor level, between $e^+ e^-$
and $\tau$ data. Indeed, Table \ref{T2}, which summarizes our simultaneous
description of $e^+ e^-$  annihilation and $\tau$ decay data, displays a fit
quality of about 80\% with quite acceptable parameter values. This may imply
as well that  ${\cal B}_{\pi \pi}$ is slightly overestimated or that some 
additional systematics affect the $1/N dN/ds$ spectrum in some experiments.

\vspace{0.5cm}

The most important aim of the present  work was to study the effects of a global fit
on the estimation of hadronic contributions to  $a_\mu$, the muon $g-2$, from the
various thesholds up to 1 GeV. For this purpose, our treatment of statistical and systematic
 errors accounted as closely as possible for the information provided 
for each data sample. 

  One first  has  checked that the fit performs as one 
  could expect on individual $e^+ e^- \ra \pi^+ \pi^-$ data samples. Comparing
  our fit results with individual experimental information 
  \cite{CMD2-1995corr,CMD2-1998-1,SND-1998}   about the contribution   
of the reference region $(0.630,~0.958)$ GeV to $a_\mu(\pi^+ \pi^-)$  is already
  interesting~: Table \ref{T3} shows the information returned by the fits is 
 in good agreement  
 with the corresponding experimental  information. Performing the global fit with
 these  data samples, where systematics have certainly been considered  with
 great care,  provides quite reasonable central values for the individual
 estimates of $a_\mu(\pi^+ \pi^-)$ and shrinked uncertainties~;
 this shrinking becomes noticeable when fitting simultaneously 
 all $e^+ e^-$ data from \cite{CMD2-1995corr,CMD2-1998-1,SND-1998}~: 
 The uncertainty is reduced from 3.02 to 1.72
 in units of $10^{-10}$, {\it  i.e.} better than a 40 \% improvement. 
 
 The next step was to include the older  $e^+ e^- \ra \pi^+ \pi^-$ data samples 
 \cite{Barkov,DM1}, where systematics are certainly not as well controlled as in 
 \cite{CMD2-1995corr,CMD2-1998-1,SND-1998}. One thus gets a shift 
 $\Delta a_\mu(\pi^+ \pi^-)\simeq -1.9$  (in units of $10^{-10}$) while
 leaving the uncertainty nearly unchanged. Including also  all the 
 $e^+ e^- \ra (\pi^0/\eta) \gamma$ and $e^+ e^- \ra \pi^0 \pi^+ \pi^-$ 
 data samples does not produce significant modifications, showing that
 the effects of systematics when combining data samples of various qualities
 may prevent improvements. However, the corresponding combined fit
does not degrade the information~; it rather allows to check the 
stability of the estimate.
 The final uncertainty improvement is at the level of 46 \%.
 
 At this point where all information on $e^+ e^- $ annihilations -- except for
 KLOE ISR data -- are considered, one ends  up with~:
 $$ a_\mu(\pi^+ \pi^-;~0.630 \div 0.958 ) = 359.31 \pm 1.62$$
(in units of $10^{-10}$) where the error combines systematic and statistical 
uncertainties and accounts for the sample--to--sample correlations. 
The probability of the underlying fit to the data is 98.4\%.
One has shown that these high probability values are a normal consequence
of  too conservative estimates of systematics into given data sets. This
has been shown not to  hide bad descriptions of some data subsets.

Including KLOE data does not modify the picture, except for  the probabilities which
may become very low, reflecting the intrinsic large minimum $\chi^2$ for this data set. 

Adding the various $\tau$ data to the set of  $e^+ e^- $ data samples, is also worth
being  made stepwise, in order to substantiate problems and specific
properties of each of the A (ALEPH), B (BELLE) and C (CLEO) data sets.
We have first compared (see Table \ref{T3b}) our reconstructed values
for $a_\mu$ with the $\tau$ based estimates of \cite{DavierHoecker}
for the reference region $0.630 \div 0.958$ GeV \cite{ZhangPriv}.
The central values are found in agreement while the uncertainties
are importantly shrinked. This is clearly an effect of our global fit/modelling.
One should also note that the parameters provided by the simultaneous fit 
to  A, B, C and the $e^+ e^- $ data provides a value for $ a_\mu$
larger than for each of A, B and C separately. Interestingly, this property is
also exhibited by the combined experimental estimate proposed by
the authors of \cite{DavierHoecker,ZhangPriv}.

In this case, we show that introducing the  A, B and C lineshapes inside
our fit procedure allows to confirm the value for $ a_\mu(\pi^+ \pi^-;~0.630 \div 0.958 )$
derived using only $e^+ e^- $ data and some more shrinking of its uncertainty~:
$$ a_\mu(\pi^+ \pi^-;~0.630 \div 0.958 ) = [359.62 \pm 1.51] ~10^{-10}~.$$
Using additionally the KLOE data leads to~:
$$ a_\mu(\pi^+ \pi^-;~0.630 \div 0.958 ) = [358.52 \pm 1.32] ~10^{-10}~.$$
Both results correspond to good probabilities of the underlying 
form factor fits.

However, our study leads us to conclude that some distortions of the lineshape,
different for A and B/C, are at work  which still need to be understood. Whether
one is faced here with  an incomplete IB distortion modelling, with unaccounted for 
systematics, or with some external physical effect, is an open issue. Until this 
issue is clarified,  
going much beyond the $\tau$ lineshapes for $ a_\mu(\pi^+ \pi^-)$ estimates
 looks hazardous.
 
Our results concerning $ a_\mu(\pi^+ \pi^-;~0.630 \div 0.958 )$ are summarized
in Figure \ref{Compilation}. This clearly illustrates that shape distortions and
systematics are not completely understood, even if at the form factor level
the picture is more optimistic~; this shows that $a_\mu$ estimates are a more
sensitive probe to differences than fit probabilities of the underlying form factors.

 Our tendency for now is to prefer providing our final results 
concerning the various contributions to $a_\mu$ from thresholds
to 1 GeV, considering all NSK $e^+e^-$ data samples together
with $\tau$ lineshape data. The KLOE data set does not seem
to modify the picture unreasonably. This information
is the matter of Table \ref{T6}. 

\vspace{0.5cm}

One may also conclude that the global fit method performs as could
be expected and is  a useful tool in order to examine the consistency
of various data sets covering various processes related by physics.
As a tool, it also permits improved estimates of the various
hadronic contributions to the muon $g-2$. 

For this purpose, one should stress 
that  better data, with better controlled systematics in all decay channels, 
and not only for the $\pi^+ \pi^-$ final state, may be valuable. Indeed, the 
physics correlations
between the various final states in $e^+ e^-$ annihilations may conspire 
(better than presently) with each other in order to provide improved values 
 for all contributions. 

Theoretical  developments on 
the inclusion of scalar mesons and higher mass vector mesons
within VMD--like frameworks  are also desirable, as this could  
increase the underlying physics correlations between the various
final states accessible in $e^+e^-$  annihilations. Finally, 
understanding the issue raised by \cite{BGPter} about the 
coupling of the $\phi$ meson to $K \overline{K}$ pairs, may well be
an important task in order to estimate reliably the $\phi$ region contribution
to the muon $g-2$. 

\section*{Acknowledgments}
\indent \indent
We are indebted to H. Hayashii, Nara Women's University, Nara, Japan,
 for having provided us with the BELLE data sample and for information 
concerning the BELLE Collaboration fits. We also thank
Z. Zhang, LAL Orsay, to have provided us with some interesting unpublished 
results quoted in the text.
Finally, M.B. gratefully acknowledges several useful discussions and 
mail exchanges with 
S. Eidelman, Budker Institute, Novosibirsk, Russia.
                   \bibliographystyle{h-physrev}

                     \bibliography{vmd2}

\end{document}